\documentclass[fleqn,usenatbib]{mnras}

\usepackage[T1]{fontenc}
\usepackage{ae,aecompl}

\usepackage{newtxtext,newtxmath}
\usepackage{graphicx}	% Including figure files
\newcommand{\kms}{\,km\,s$^{-1}$} % kilometres per second
\title[A new scenario for Abell 1644]{Revising the merger scenario of the galaxy cluster Abell 1644: a new gas poor structure discovered by weak gravitational lensing}

\author[Monteiro-Oliveira et al.] %Rogério Monteiro de Oliveira
 {R.~Monteiro-Oliveira,$^{1}$\thanks{E-mail: rogerionline@gmail.com}
  L.~Doubrawa,$^{1,2}$ 
  R. E. G.~Machado,$^2$
  G. B.~Lima Neto,$^1$
  \newauthor % starts a new line in the author environment
  M.~Castejon$^1$
   and
  E. S.~Cypriano$^1$\\
  $^1$Universidade de S\~ao Paulo, Inst. de Astronomia, Geof\'isica e Ci\^encias Atmosf\'ericas, Depto. de Astronomia, R. do Mat\~ao 1226, 05508-090 S\~ao Paulo, Brazil\\
  $^2$Universidade Tecnol\'ogica Federal do Paran\'a, Rua Sete de Setembro 3165, 80230-901 Curitiba, Brazil\\
 }

\date{Accepted XXX, Received YYY in original form ZZZ}

\pubyear{2020}

\begin{document}
\label{firstpage}
\pagerange{\pageref{firstpage}--\pageref{lastpage}}
\maketitle

\begin{abstract}
The galaxy cluster Abell 1644 ($\bar{z}=0.047$) is known for its remarkable spiral-like X-ray emission. It was previously identified as a bimodal system, comprising the subclusters, A1644S and A1644N, each one centred on a giant elliptical galaxy. In this work, we present a comprehensive study of this system, including new weak lensing and dynamical data and analysis plus a tailor-made hydrodynamical simulation. 
The lensing and galaxy density maps showed a structure in the North that could not be seen on the X-ray images. We, therefore, rename the previously known northern halo as A1644N1 and the new one as A1644N2. Our lensing data suggest that those have fairly similar masses: $M_{200}^{\rm N1}=0.90_{-0.85}^{+0.45} \times10^{14}$  and $M_{200}^{\rm N2}=0.76_{-0.75}^{+0.37} \times10^{14}$ M$_\odot$, whereas the southern structure is the main one: $M_{200}^{\rm S}=1.90_{-1.28}^{+0.89}\times 10^{14}$ M$_\odot$. 
Based on the simulations, fed by the observational data, we propose a scenario where the remarkable X-ray characteristics in the system are the result of a collision between A1644S and A1644N2 that happened $\sim$1.6 Gyr ago. Currently, those systems should be heading to a new encounter, after reaching their maximum separation.
\end{abstract}


%%%%%%%%%%%%%%%%%%%%%%%%%%%%%%%%%%%%%%%%%%%%%%%%%%

%%%%%%%%%%%%%%%%% BODY OF PAPER %%%%%%%%%%%%%%%%%%

\section{Introduction}
\label{sec:intro}

%<--General introduction to large scale structure
The formation of galaxy clusters through the merger of smaller structures is one of the most energetic events known in the Universe, involving an amount of energy of about $10^{64}$~erg \citep{sarazin04}.
%% comparable only with the Big Bang. Acho que o Big Bang foi BEM mais do que isto (nao e' realmente "comparavel")
However, since this process runs over a long period of time \citep[$1-4$ Gyr; e.g.][]{Machado+2015}, it does not constitute a cataclysmic event. Nevertheless, merging clusters are a fundamental cornerstone of large-scale structure formation as predicted by our standard cosmological scenario \citep[e.g.,][]{lacey93,millennium05} and have an ubiquitous signature, the presence of substructure on clusters of galaxies \citep[e.g.,][]{beers82,Jones99,AndradeSantos2012}.

%<--Something more specific
In spite of the merger process among galaxy clusters being spread over a long period of time, the moments immediately after the pericentric passage are particularly interesting. It has been seen that each cluster component -- dark matter, intracluster gas and galaxies -- behaves in particular ways during the collision, and a spatial detachment between them can be observed just after the pericentric passage \citep{markevitch04}. 
It has been claimed that a detachment between galaxies and dark matter would be an observational signature of self-interaction of the dark matter and can provide an astrophysical tool for computing its cross section \citep[$\sigma/m$; e.g.][]{harvey14,harvey15}. Whether or not such a detachment have been actually detected on data is a matter of debate  \citep{Monteiro-Oliveira17a, Wittman17}.

%<--What about the non-dissociative mergers?
Even when the merger does not belong to the dissociative class \citep{dawson}, interesting observational features can arise from cluster interactions. For instance, the gravitational perturbation due to an off-axis encounter with a smaller structure may stir the cool gas from the cluster core. This mechanism is known as gas sloshing \citep{Markevitch01, markevitch_viki07} and may give rise to a spiral of low-entropy cool gas that stems from the cluster core reaching out to as much as a few hundred kpc. Such spirals, which are detectable as an excess in X-ray emission, are not so rare. Based on deep {\it Chandra} observations, \citet{Lagana2010} showed that this phenomenon may indeed be quite common in the nearby universe. One prominent example of this kind of object is Abell 2052 \citep{Blanton2011}, which has also been modelled by hydrodynamical $N$-body simulations as an off-axis collision \citep{Machado2015}. In the Perseus cluster, detailed features of the sloshing cold front have also been studied with dedicated simulations \citep{Walker2018}. More recently, a novel technique has allowed measurements of bulk flows of the intracluster medium (ICM), offering direct evidence of gas sloshing in the potential wells of Coma and Perseus \citep{Sanders2020}.

\begin{figure*}
\begin{center}
\includegraphics[width=\textwidth, angle=0]{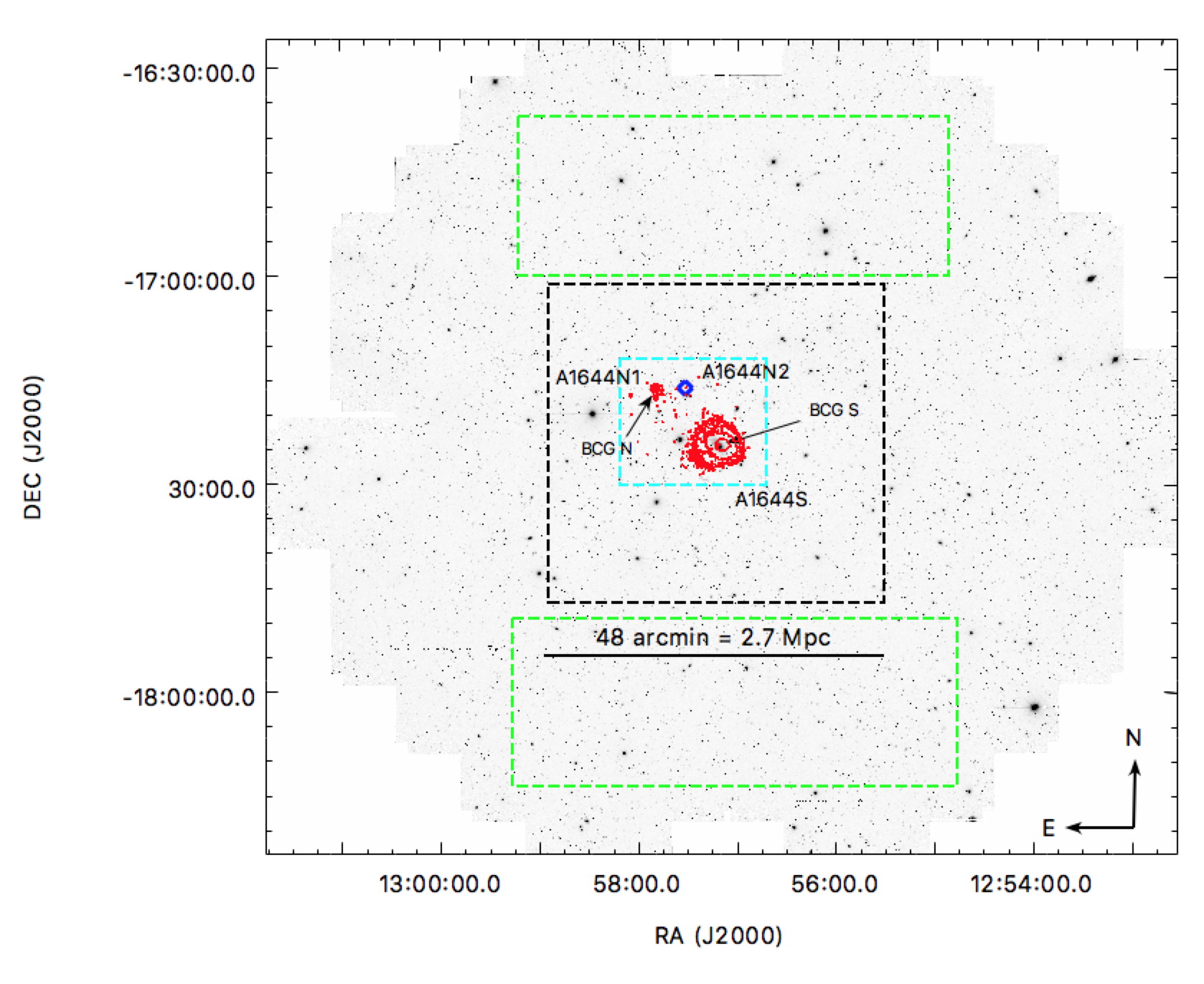}
\caption[]{Optical image of the A1644 field in the $r'$ band, taken with the Dark Energy Camera at Victor Blanco telescope. The red contours correspond the \textit{Chandra} X-ray image. The X-ray emission is bimodal and each peak coincides with the position of a giant elliptical galaxy (BCG~S and BCG~N), located at the centre of the main (A1644S) and northern (A1644N1) structures. The new substructure found by weak lensing (Sec.~\ref{sec:wl}) is labelled as A1644N2 (blue diamond).
 Green and cyan dashed boxes delimit the regions used the identification of the cluster red sequence galaxies (Sec.~\ref{sec:photo}). The black dashed line box encloses the region of interest (48 arcmin$^2$), which we will analyse throughout the paper.} 
\label{fig:field}
\end{center}
\end{figure*}

%<---Previous X-ray analysis
In this work, we will devote our attention to the nearby galaxy cluster Abell 1644 \citep[hereafter A1644; $\bar{z}=0.047,$][]{Tustin01}, which we show in Fig.~\ref{fig:field}.
An {\it Einstein} X-ray image by \cite{Jones99} showed A1644 as a bimodal structure. \citet{Johnson10} named the components after their positions as A1644S, the main (southern) structure, and  A1644N on the North side. Both have their X-ray peaks spatially coincident with a giant elliptical galaxy or their respective brightest cluster galaxies (BCGs). \cite{Reiprich04}, using XMM-{\it Newton} data, argued that we are seeing the system after an off-axis collision of those substructures. In Table~\ref{tab:mass.comp} we present a compilation of estimated masses for A1644 available in the literature.

%<--Figure: A1644 field
%

%<--Table: Previous mass measurements
\begin{table}
%\begin{center}
\caption[]{A1644 mass according to several methods. RVD stands for radial velocity dispersion; WGL for weak gravitational lensing.}
\setlength{\tabcolsep}{2pt}\centering
\begin{tabular}{lccc}
\hline \hline 
Region & $M_{200}$ & Method & Reference \\%[5pt]
       & $(10^{14} h_{70}^{-1}$ M$_{\odot})$ & & \\
%&($10^{14}$ M$_{\odot}$) $h_{70}^{-1}$ & & \\[5pt]
\hline 
Overall & $6.0$  &   RVD plus $T_X$   & \cite{Ettori97}\\[5pt]
Overall & $4.50_{+0.80}^{-0.75}$      &   RVD   & \cite{girardi98}\\[5pt]
Overall & $6.9\pm1.0$                 & Caustic$^\dagger$ & \cite{Tustin01}\\[5pt]
Overall & $10.40_{+1.01}^{-0.94}$     & RVD -- NFW  & \cite{Lopes18}\\[5pt]
Overall$^\diamondsuit$ & $3.99_{-1.59}^{+1.39}$ & WGL -- NFW   & This work (Sec.~\ref{sec:wl})\\%[5pt]
\hline
A1644N1 & $3.8\pm0.6$    & $M_{500}\times T_{\rm X}$    & \cite{Johnson10}\\[5pt]
A1644N1 & $0.90_{-0.85}^{+0.45}$  & WGL -- NFW   & This work (Sec.~\ref{sec:wl})\\%[5pt]
\hline
A1644S  & $4.6\pm0.6$    & $M_{500}\times T_{\rm X}$    & \cite{Johnson10}\\[5pt]
A1644S & $1.90_{-1.28}^{+0.89}$  & WGL -- NFW   & This work (Sec.~\ref{sec:wl})\\%[5pt]
\hline
A1644N2$^\star$ & $0.76_{-0.75}^{+0.37}$  & WGL -- NFW   & This work (Sec.~\ref{sec:wl})\\
\hline \hline 
\multicolumn{4}{l}{$^\dagger$ Inside a radius $R = 2.4$ Mpc}\\
\multicolumn{4}{l}{$^\diamondsuit$ This value corresponds to the sum of all mass clumps identified by us}\\
\multicolumn{4}{l}{~~~as A1644 members.}\\
\multicolumn{2}{l}{$^\star$ New structure found in this work.}\\
\end{tabular}
\label{tab:mass.comp}
%\end{center}
\end{table}

%<--Optical
Regarding the optical components, the substructures of A1644 are somewhat less clearly visible. Whereas \cite{ds} found evidence for substructure in A1644 based on the analysis of 92 redshifts, \cite{Girardi97} classified A1644 as a single system, based on an analysis combining both galaxy positions and redshifts. A similar conclusion was proposed by \cite{Tustin01} who measured redshifts for 144 cluster members. They found a rather larger than expected  radial velocity dispersion, $\sigma\sim 1000$ km~s$^{-1}$, which suggested a system out of the equilibrium state \citep[e.g.][]{Monteiro-Oliveira18, Pandge19}. However, the authors found no evidence for the presence of substructure in this cluster, in contrast with the scenario suggested by X-ray observations.

%<--Johnson et al. 2010
High spatial resolution observations with the {\it Chandra} satellite analysed by \cite{Johnson10} provided straightforward arguments in favour of the post-collision state of A1644. Notably, the authors detected the presence of a cold front \citep{markevitch_viki07} at the edge of the Southern subcluster with a  spiral morphology. They proposed a merger scenario where an off-axis passage of A1644N by A1644S was responsible for pushing the gas of the latter out the bottom of its gravitational potential well. This gas was driven into an oscillating motion (sloshing) which generated the observed spiral-like structure.  Based on a comparison with generic hydrodynamical simulations \citep{Ascasibar06}, they estimated that the pericentric passage between A1644S and A1644N happened about 700 Myr ago.

%<--What is lacking in the literature?
In spite of the wealth of analysis already devoted to A1644, a mapping of the total mass distribution is still lacking in the literature. One plausible reason is that weak lensing studies of such low-redshift systems are challenging due to the intrinsic low signal \cite[e.g.][]{cypriano01,cypriano04}. Nevertheless, we have successfully recovered the mass distribution of a galaxy cluster at a similar redshift \citep[Abell 3376;][]{Monteiro-Oliveira17b}. 

%<--This work
This paper presents the first weak lensing study fully dedicated to A1644. From deep and large field-of-view multiband images ($g'$, $r'$, and $i'$), we recovered the mass distribution of the cluster. Complementarily, the redshift catalogue available in the literature allowed us to revisit the cluster dynamics in order to solve the inconsistency between the distinct cluster morphologies proposed previously by either X-ray or radial velocity observations. From these combined analyses (weak gravitational lensing plus cluster dynamics), we were able to characterize dynamically each component of the system, including substructures that were not identified in previous studies. To go beyond our present understanding of this cluster's dynamical history, we have performed a tailor-made hydrodynamical simulation in order to describe the timeline of the collision. This tool is fundamental to link the observational findings of this paper and the remarkable X-ray features seen in A1644 to the process of large-scale structure formation. 

%<--Organization of the paper
This paper is organized in the following way. In Section~\ref{sec:photo} we perform the photometric analysis. The reconstruction of the mass field through the weak gravitational lensing is introduced in Section~\ref{sec:wl}. The dynamical view, based on the redshift catalogue is presented next, in Section~\ref{sec:dynamical}. The set-up as well as the results of our hydrodynamical simulation showing the best model for the collisions that occurred in A1644 are in Section~\ref{sec:hidro}. Finally, our main findings are discussed in Section~\ref{sec:discussion} and summarised in Section~\ref{sec:summary}.

%<--Cosmology
Throughout this paper we adopt the standard $\Lambda$CDM cosmology, represented by $\Omega_m=0.27$, $\Omega_\Lambda=0.73$, $\Omega_k = 0$ and $h = 0.7$. At the mean cluster redshift of $z = 0.047$, we then have a plate scale of 1 arcsec equal 0.926 kpc
%% , the age of the Universe 13.9 Gyr (para que isto??)
and an angular diameter distance of 191.1 Mpc \citep{CosmoCalc}.

\section{Photometric analysis}
\label{sec:photo}

\subsection{Imaging data}
\label{sec:data}

%<--Origin of the data 
The present imaging data were obtained on February 1, 2014 with the Dark Energy Camera (DECam) mounted at the Victor Blanco 4m-telescope (Proposal ID: 2013B-0627 within the SOAR Telescope time
exchange program; PI: Gast\~ao Lima Neto). The observational details are shown in Table~\ref{tab:imaging}. 

%<--Imaging details
\begin{table}
\begin{center}
\caption[]{Characteristics of our imaging data observed at Dark Energy Camera. The deepest $r'$ band was chosen as the basis for our weak lensing analysis.}
\begin{tabular}{lccc}
\hline
\hline
Band & Total exposure (h) & Mean air mass & Seeing (arcsec) \\
\hline
$g'$  & 0.5  &  1.20 & 1.21\\
$r'$   & 1.0  &  1.07 &  1.09 \\
$i'$   & 0.4  &  1.12 & 1.08 \\
\hline
\hline
\end{tabular}
\label{tab:imaging}
\end{center}
\end{table}

%<--Data reduction and problems 

We used standard procedures \citep{valdes14} for image reduction/combination and astrometric calibration. The latter resulted in a positional rms for  \citet{adelman-mccarthy} catalogue stars of $0.6\pm0.5$ arcsec, which is more than enough for the purposes of the current analysis.

Observations were made under non-photometric conditions. This is acceptable as accurate absolute photometric calibration is not a strong requirement for lensing, which is more concerned with galaxy shapes and positions.  In order to get an approximate calibration we used average values of the airmass extinction coefficients provided by the observatory staff (Walker, private communication), which resulted in the following zero point magnitudes: $g_0=31.31\pm0.05$, $r_0=31.53\pm0.05$ and $i_0=31.73\pm0.04$.

%<--Photometric catalogues 
We used  {\sc SExtractor} in double image mode to create object catalogues. Detections were made in the deeper  $r'$ image. Galaxies were then selected with two criteria: 
 (i.) $r'\leq 19.0$  {\sc CLASS\_STAR} index lower than 0.8 and (ii.) $r'> 19.0$  with their full width at half-maximum (FWHM) greater than $1.3$ arcsec, a value  $0.2$ arcsec above the seeing to ensure the selection of well-resolved objects.

\subsection{Identification of the red sequence cluster galaxies}
\label{sec:red}

%<--About the red-sequence galaxies
Galaxies correspond only to $\sim$ 5\% of the total cluster mass but their correlation in the projected space is one of the few ways to trace a cluster in the optical band. Another useful tracer of the red cluster members is their well-known correlation in the colour-colour map \citep[e.g.][]{med10, Medezinski18}. We identified the A1644 red-sequence {\it locus} on an $(r'-i') \times (g'-r')$ diagram by applying a statistical subtraction method \citep[more details in ][]{Monteiro-Oliveira17b}.

\begin{figure}
\begin{center}
\includegraphics[width=\columnwidth, angle=0]{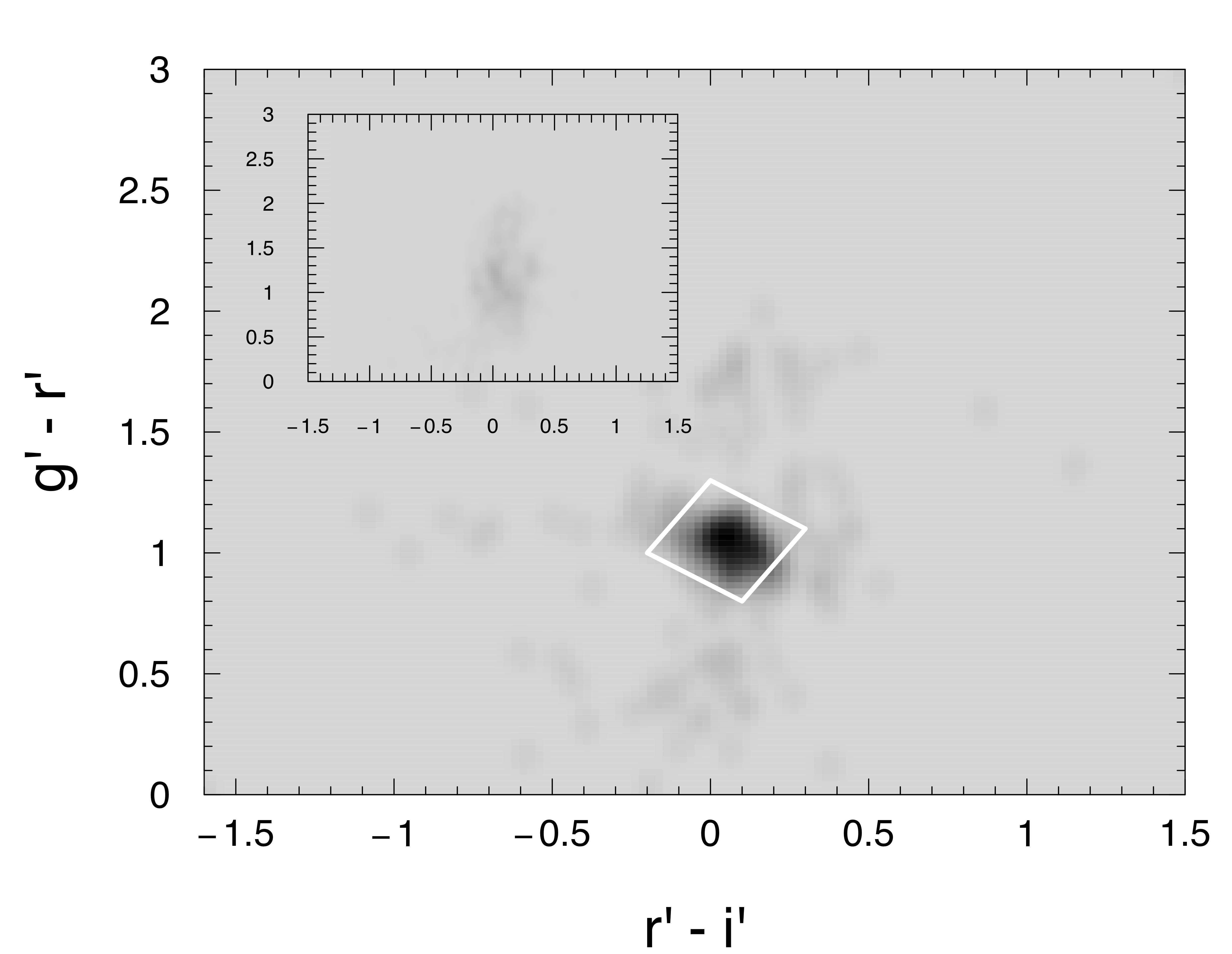}
\caption[]{Colour-colour diagram of the galaxies located at the innermost region of the cluster field (cyan box of Fig.~\ref{fig:field}). Due to the fact that they present homogeneous photometric properties, the red cluster member galaxies form a density peak in this space, thus defining the cluster member {\it locus}. In this plot, galaxies are brighter than $r'=19.5$, considering the magnitude limit for selection of cluster members.  The inset panel presents the same plot with the same linear grey-scale, but for the galaxies in the control areas (green boxes in Fig.~\ref{fig:field}).} 
\label{fig:cc}
\end{center}
\end{figure}

We used the galaxies within the green boxes in  Fig.~\ref{fig:field} as a representation of the field population and the ones inside the cyan box as field plus cluster populations. This central region presents a galaxy  density excess in relation to the control (field) ones up to magnitude $r'=19.5$, which we than took as a faint limit for A1644 red-sequence galaxies. In Fig.~\ref{fig:cc} we show this density excess in the colour-colour space. The region with the highest excess corresponds to the locus of the A1644 early-type galaxies.

\begin{figure}
 \begin{center}
\includegraphics[width=\columnwidth, angle=0]{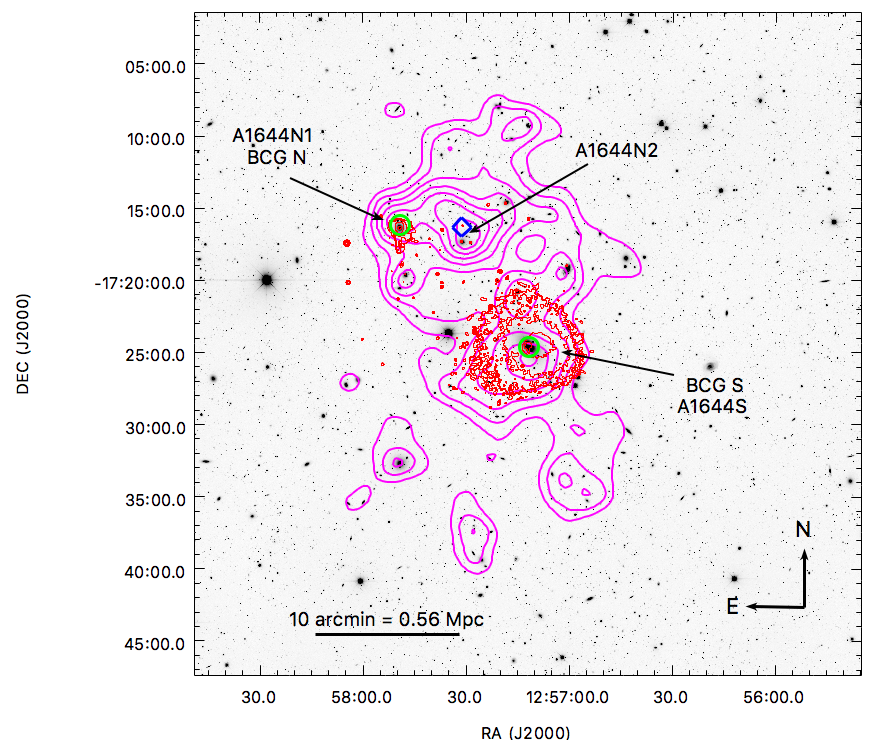}
\caption[]{Optical image overlaid with contours of the cluster member luminosity map  (magenta) and of the X-ray emission (red). Both BCG\,S and BCG\,N (green circles) lie in local peaks of the luminosity map. Furthermore, this shows a third relevant concentration west of A1644N1, within the \textit{Chandra} field but with no visible X-ray emission. We named this substructure as A1644N2. As we show in the text (Sec.~\ref{sec:rec.mass}), this galaxy concentration coincides with the position of a dark matter clump (blue diamond). In addition, we found other less significant clumps surrounding the central region. This image is of the black box area defined in Fig.~\ref{fig:field}.} 
\label{fig:density}
\end{center}
\end{figure}

There is a total of 613 galaxies on this colour-colour locus over the entire optical image. We call those `photometric members', yet knowing that this sample is not free of contamination and that it does not include blue members. Their $r'$ projected galaxy luminosity distribution is presented in Fig.~\ref{fig:density}. This `luminosity map' is a 30~arcsec pixel-based image done by smoothing each galaxy by a bidimensional, circular Plummer profile (or an index 5 polytrope) kernel with a core radius of 1~arcmin, following \citet{OMill2015} and \citet{Machado2015}.

%<--Figure: Colour-colour diagram

%12:57:31.1135
%-17:16:21.413
%<--Figure: red-sequence projected density

%<--Comments about the members distribution.

Although we will perform a more quantitative analysis of the red cluster galaxy luminosity peaks after the lensing analysis in Sec. \ref{sec:rec.mass}, there are obvious features in Fig. \ref{fig:density} that should be pointed out. There X-ray maxima (A1644S and A1644N1) have their luminosity counterparts. There is one very prominent peak in the red member luminosity map with no X-ray counterpart. We label this substructure as A1644N2.

\section{Weak lensing analysis}
\label{sec:wl}

%<--Short introduction.
We used weak lensing techniques to recover the projected mass distribution on the A1644 field as estimate masses of its substructures, following the procedures we used in the study of other merging clusters \citet{Monteiro-Oliveira17a,Monteiro-Oliveira17b,Monteiro-Oliveira18}. 
Below we describe the data characteristics, treatment, and modelling. For a review of the fundamentals of lensing, we refer the reader to excellent reviews that can be found in the literature, such as \cite{mellier99}, \cite{schneider05}, and \cite{schneider06}.

\subsection{Source selection and shape measurement}
\label{sec:back}

%<--The importance of done a good source selection.
The background or source galaxies images constitute the raw material for weak lensing studies because those respond to the gravitational lensing effect caused by the galaxy cluster mass. This effect manifests as a slight shape distortion on the images of the aforementioned galaxies. Therefore, a careful selection of source galaxies is crucial for  a successful weak lensing analysis. 

%<--How we have selected source galaxies.
A1644 is located at a very low redshift ($\bar{z}=0.047$, see Sec.~\ref{sec:dynamical}), therefore we expect most galaxies in our image to be in the cluster background. In fact, for a galaxy cluster at a similar distance, we found that foreground galaxies correspond to less than 0.2\% of the CFHTLS\footnote{Canada-France-Hawaii Telescope Legacy Survey; \url{http://www.cfht.hawaii.edu/Science/CFHTLS/}} catalogue \citep{Monteiro-Oliveira17b}. This prompted us to consider as source all galaxies fainter  than $r' = 21$ and  not in the red cluster locus on the colour-colour space (Fig.~\ref{fig:cc}). Remembering that we could not detect an excess associated with clusters galaxies fainter than  $r' = 19$, we should have a very complete and uncontaminated source sample. 

To estimate the average surface critical density, that depends on both the cluster and source redshifts, we applied the same cuts on the CFHTLS catalogue, with photometric redshifts, and found $\Sigma_{\rm cr}=9.6\times10^9$ M$_\odot$ kpc$^{-2}$.

%<--Measuring the shapes
The next step is to map the point spread function (PSF), which affects all image shapes across the $r'$ band image used for shape measurements. To do that, we identified bright unsaturated stars spread across the field, split them into nine frames and modelled their shapes with  the Bayesian code {\sc im2shape} \citep{im2shape}. Besides modelling individual objects as a sum of Gaussians with an elliptical basis, this code also deconvolves the effects of the local PSF.

\begin{figure*}
 \begin{center}
\includegraphics[width=1.0\textwidth,angle=0]{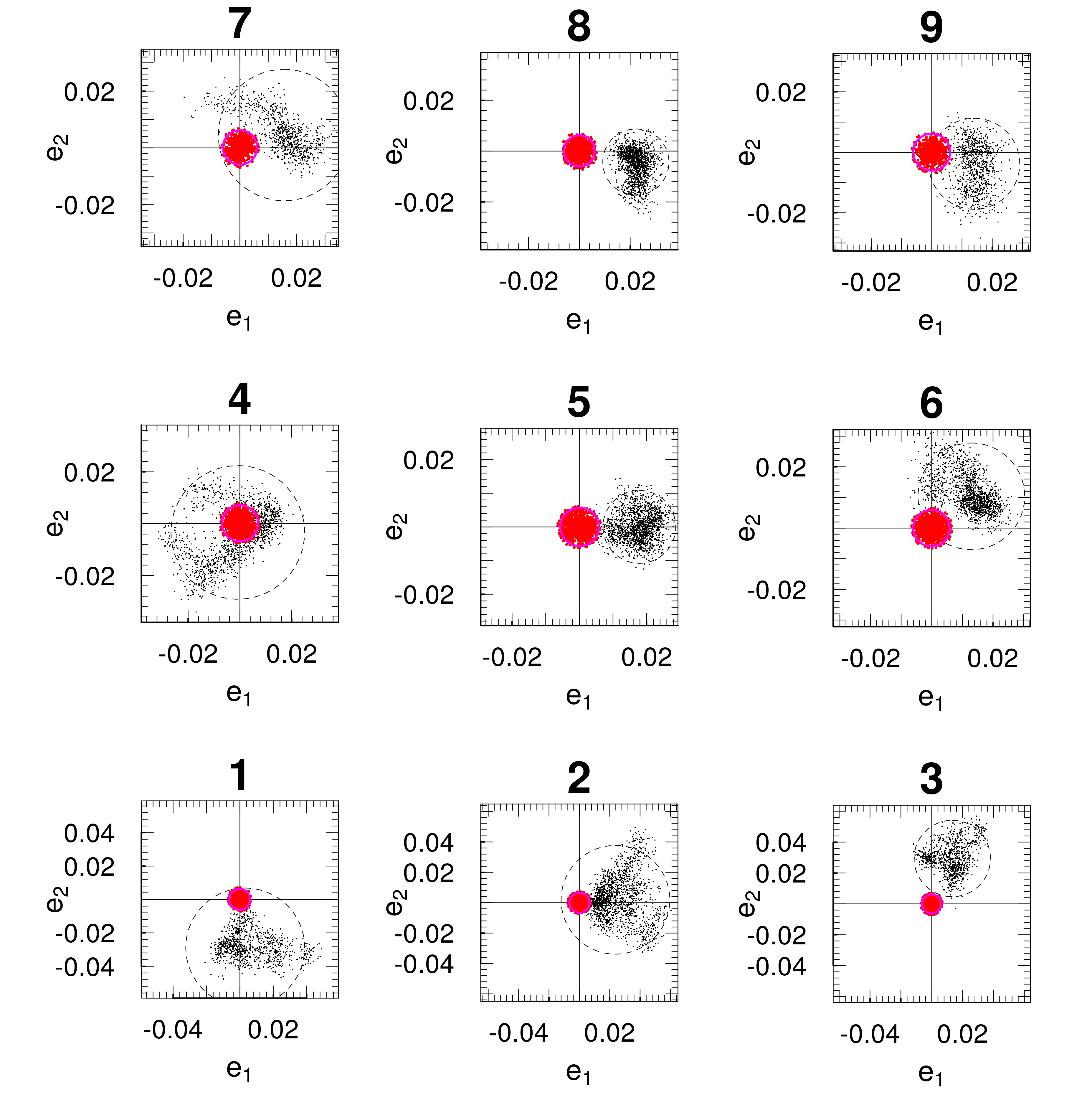}
\caption[]{PSF ellipticity. The black dots are the raw components $e_1$ and $e_2$ obtained from the stars and the red ones the residuals between those and the PSF elliticity map for the 9 segments that composes a full DECam image. Dashed circles enclose 95\% of the data. The averaged residuals are $~10^{-5}$ for $e_1$ and $e_2$ with a standard deviation of 0.003 for both.}
\label{fig:psf.stars}
\end{center}
\end{figure*}

For the PSF mapping, the star profiles are modelled as single Gaussians and no PSF deconvolution is done. The main results of this process are the ellipticity  components $e_1$, $e_2$, and a size parameter, related to the seeing. Then, the discrete sample of parameters was spatially interpolated across the field with the thin plate spline regression \citep{fields} built-in R environment \citep{R} to create a continuous function.
Stellar data with discrepant values were removed in 3 iterative steps; at each step, objects with the 10 per cent largest absolute residuals were discarded. The measured stellar ellipticities and respective residuals after the spatial interpolation are presented in Fig.~\ref{fig:psf.stars}.

%<--Figure: PSF measurement

%<---Coments on PSF modelling
The averaged ellipticity of all field was $\langle e_1 \rangle=0.014$ with $\sigma_{e_1}=0.010$ and  $\langle e_2 \rangle=0.001$ with $\sigma_{e_2}=0.010$. As stars are points sources for all practical aspects all deviations from null values are due to the PSF effect. As illustrated by Fig.~\ref{fig:psf.stars}, our model performed a good match with the data as shown by the very small averaged residual ($10^{-5}$) and standard deviation ($0.003$).

Given the PSF field at each point of the image, we used {\sc im2shape} over galaxies, modelling them as two Gaussians with the same elliptical base. Now the software produces PSF-free ellipticities and respective uncertainties.  The latter are used as a quality criterion. Objects with $\sigma_e>2$ or those having evidence of blending were removed from the source sample.

\subsection{The projected mass distribution}
\label{sec:rec.mass}

\subsubsection{The signal-to-noise shear map}

\begin{figure*}
\begin{center}
\includegraphics[width=1.0\textwidth, angle=0]{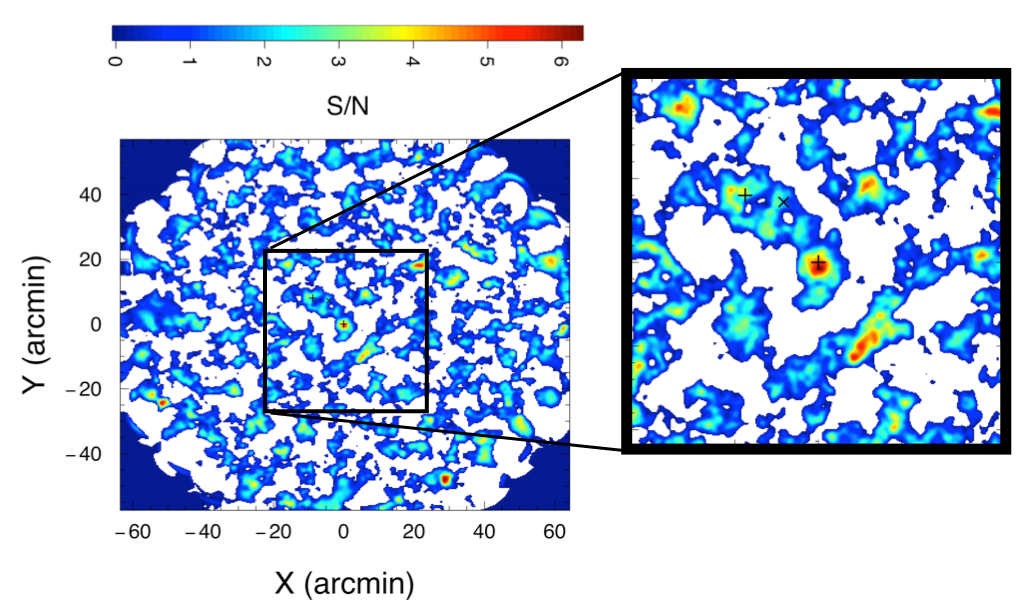}
\caption[]{Signal-to-noise map obtained through the mass aperture statistic highlighting A1644 central region (Fig.~\ref{fig:field}), according to the cluster member luminosity distribution. White regions correspond to negative S/N. The position of the two BCGs and the galaxy clump A1644N2 are marked with ``+'' and ``$\times$'', respectively. The field is densely populated by mass peaks, most of them, however, due to the background large-scale structure.}
\label{fig:sn}
\end{center}
\end{figure*}

The ellipticities of the galaxies can be seen as noisy probes of the shear field, as each galaxy has its own non-lensing related ellipticity. The shear field can then, through techniques, be turned into a convergence field or a mass map. Here we created a signal-to-noise ratio (S/N) map of structures by averaging the tangential ellipticity $e_+$ in relation of a grid of points, for the $N_{\theta_0}$ galaxies inside a radius $\theta_0$, as prescribed by the mass aperture statistic \citep{schneider96},
\begin{equation}%\dfrac{M_{\rm aper}}{\sigma_{\rm aper}}=
{\rm S/N}=\dfrac{\sqrt{2}}{\sigma_e^2}\ \dfrac{\sum_{i=1}^{N_{\theta_0}} e_{+_i}(\theta_i) Q_{\rm NFW}(\theta_i,\theta_0)}{\left[ \sum_{i=1}^{N_{\theta_0}} Q_{\rm NFW}^2(\theta_i,\theta_0)\right ]^{1/2}} \ \mbox{,}
\label{eq:SN}
\end{equation} 
%<--S/N
where $\theta_i$ is the position of the $i^{\rm th}$ source galaxy and $\sigma_e$ is the quadratic sum of the measured error and the intrinsic ellipticity uncertainty, estimated as $0.35$ for our data. We adopted the NFW filter \citep{schirmer04},
\begin{multline}
\noindent Q_{\rm NFW}(\theta_i,\theta_0)=[1+e^{a-b\chi(\theta_i,\theta_0)}+e^{-c+d\chi(\theta_i,\theta_0)}]^{-1} \times  \\
\dfrac{\tanh [\chi(\theta_i,\theta_0)/\chi_c]}{\pi\theta_0^2[\chi(\theta_i,\theta_0)/\chi_c]}\mbox{,}
\label{eq:nfw.filter}
\end{multline}
with $\chi=\theta_i/\theta_0$, which describes approximately an NFW shear profile. \cite{hetterscheidt05} suggest, as optimized parameters for halo detection, $a=6$, $b=150$, $c=47$, $d=50$, and $\chi_c=0.15$, which we adopted. We also considered the radius $\theta_0=8$ arcmin. The resulting S/N map can be seen in Fig.~\ref{fig:sn}.

%<--Figure: A1644 S/N

%<--Comments about S/N

In a case like A1644, which is not a very massive cluster and is situated at a low redshift, that does not favour the lensing signal, its structures do not stand out in the shear S/N map. In such a case several smaller mass background structures plus line-of-sight superpositions \citep{Yang13,Liu16} can create similar S/N features in the map and thus the noisy appearance we see in Fig.~\ref{fig:sn}.
None the less the map shows a clear high S/N clump spatially coincident with the BCG S. Also, BCG N is superposed with an arched clump that stretches mainly in the east-west direction. Besides, there is some amount of significance halfway between the two BCGs, whose pertinence to the targeted galaxy cluster will be better investigated.

%<--LensEnt2 

\subsubsection{Convergence map}
\label{sec:kappamap}
To create a convergence ($\kappa$) or projected mass map of the field we resorted to the  maximum entropy algorithm \citep{seitz98} of the Bayesian code {\sc LensEnt2} \citep{LensEnt2}. It works by maximizing the evidence of the reconstructed mass field in relation to the data. In order to take into account the noisiness of individual measurements and prevent overfitting. {\sc LensEnt2} uses a Gaussian intrinsic correlation function (ICF). We have tested reconstructions done with a $\sigma_{\rm ICF}\in[70:210]$ arcsec range.

For each realization of a mass map, given the ICF width, we did an automatic peak search (which is described next) and compared the number of detected clumps above 1$\sigma$, 2$\sigma$, and 3$\sigma$. We found that the number of peaks detected above $3\sigma$ remained approximately the same for $\sigma_{\rm ICF} \ge 170$~arcsec ($\sim 2.8$~arcmin), which we adopted as bona fide.  The convergence  map is presented in Fig.~\ref{fig:mass.red} and the mass reconstruction characteristics are shown in Table~\ref{table:mass.charac}.

%<--Mass reconstruction characteristics
\begin{table}
\begin{center}
\caption[]{Relevant quantities for the weak lensing convergence map reconstruction.}

\begin{tabular}{l c}
\hline
\hline
$N_g$ (gal. arcmin$^{-2}$)  &  10.7\\ 
ICF FWHM  (arcmin) &  2.8 \\
$\sigma_\kappa {}^\star$  & 0.011\\
\hline
\hline
\end{tabular}
%\begin{tablenotes}
      {\small
      \item  ${}^\star$ Noise level in the convergence map.}
%\end{tablenotes}
\end{center}
\label{table:mass.charac}
\end{table}

%<--Find the cluster mass counterparts
The S/N and the convergence maps are qualitatively very similar, both showing a wealth of structures. Some common features stand out, however; in particular, the peaks \#2 and \#6 of the $\kappa$ map, which are also present in the S/N map, and can be associated with A1644S and A1644N1, respectively, which we already identified in the member luminosity and X-ray maps. The situation is not so clear for the remaining peaks and will be further investigated.

%<--Characterisation of the mass peaks
\subsubsection{Characterization of the mass peaks}

To compute the exact position of each mass clump or peak centre, we applied an automatic procedure. First, the algorithm searched for local maxima in the convergence map within a moving circular window with 2.3 arcmin radius, which allows measuring individual peak statistics (e.g. $\bar{\kappa}$) without any overlapping. 

The peak centre position is then defined as the pixel-weighted mean inside the circular region. Its significance $\nu$ is the ratio between the local maxima $\kappa_{\rm max}$ and the noise level $\sigma_\kappa$.
To measure the latter, we have performed 100 realizations of the mass distribution removing the cluster lens signal. This was done by rotating each galaxy ellipticity by a random angle [0,180[. This procedure also allowed us to identify spurious peaks caused by noise fluctuations.

The averaged noise level can be seen in Table~\ref{table:mass.charac} and a summary of the detected and expected number of spurious peaks are in Table~\ref{table:peak.detection}. As we can see, the probability to detect a spurious peak above $4\sigma_\kappa$ is almost zero. In Table~\ref{tab:masses}, we describe the most significant ($\nu>4$) peaks within the convergence map. Some of those maybe be related to  A1644 but most should be typical field features of such maps \citep{Yang13,Liu16}.
% and related to A1644.

%<--Detection features
\begin{table}
\begin{center}
\caption[]{Summary of our search for mass peaks in the convergence map.
}
\begin{tabular}{c c c }
\hline
\hline
Threshold interval & Number of & Expected number of  \\
($\sigma_\kappa$)          & detected peaks & spurious peaks \\
\hline
%%%\red{[1,2[}   & \red{1}  & $21.3\pm3.1$  \\

$[2,3[$   & 4  & $2.2\pm2.9$ \\
$[3,4[$   & 5  & $0.2\pm0.5$ \\
$\geq4$   & 18 & $0.01\pm0.10$ \\
\hline
\hline
\end{tabular}
\label{table:peak.detection}
\end{center}
\end{table}

%<--Table: modelling results
\begin{table*}
\caption[]{Statistics of the identified mass clumps. The first five columns refer to the clump ID [1], the peak centre coordinates [2--3], the peak significance $\nu=\kappa_{\rm max}/\sigma_\kappa$ [4] and the mean convergence inside a circular region [5]. The mass modelled according to our model are presented in column [6] where absent values correspond to those peaks for which our model was not able to produce reliable results (mostly because of the relatively low significance and the location close to field border). Relevant comments on each mass clump are shown in the last column [7].}
\label{tab:masses}
\begin{center}
\begin{tabular}{c c c c c c c}
\hline
\hline 
 Clump & $\alpha$ (J2000) & $\delta$ (J2000) & $\nu$ & $\bar{\kappa}$ ($<{\rm 2.3}$ arcmin) & $M_{200}$ ($10^{14}$ M$\odot$) & Comments \\
\hline
	1 & 12:55:42  & -17:06:47 &  13.0 & 0.081  & -- & Border/background \\[5pt] 
	%%%1 & 12:55:42  & -17:06:47 &  13.0 & 0.081  & $6.52_{-3.38}^{+2.54}$ & Border/background \\[5pt] 	
	2 & 12:57:12  & -17:24:47 &  10.6 & 0.071  & $1.90_{-1.28}^{+0.89}$ & A1644S\\[5pt]	
	3 & 12:56:51 & -17:34:41 &  9.9 & 0.065  & $3.37_{-2.06}^{+1.40}$  & A1644 candidate\\[5pt]
	4 & 12:58:19  & -17:06:21 &  8.7 & 0.053  & -- & Border/background \\[5pt]	
	%%%4 & 12:58:19  & -17:06:21 &  8.7 & 0.053  & $4.41_{-2.73}^{+1.81}$ & Border/background \\[5pt]		
	5 & 12:57:37  & -17:20:44  &  8.0 & 0.054   & $0.87_{-0.82}^{+0.43}$ & A1644 candidate	\\[5pt]	
	6 & 12:57:54  & -17:16:20 &  7.4 & 0.055  & $0.90_{-0.85}^{+0.45}$ & A1644N1\\[5pt]
	7 & 12:56:45  & -17:15:48 &  7.2 & 0.048  & $0.81_{-0.74}^{+0.40}$ & background \\[5pt]	
	8 &  12:55:51 & -17:24:15 &  6.4 & 0.042  & -- & Border/background\\ [5pt]
%%%	8 &  12:55:51 & -17:24:15 &  6.4 & 0.042  & $1.46_{-1.37}^{+0.70}$ & Border/background\\ [5pt]	
	9 & 12:57:31  & -17:16:21 &  6.2 & 0.046  & $0.76_{-0.75}^{+0.37}$ & A1644N2\\ [5pt]
	10 & 12:57:16 & -17:10:51 & 5.6 & 0.036  & $0.56_{-0.55}^{+0.28}$ & A1644 candidate\\ [5pt]
	
	\hline
	11 &12:58:23  & -17:30:13  &  5.4 & 0.034  &  -- & Border/background  \\ [5pt]	
	12 &  12:55:46 & -17:29:14 &  4.6 & 0.035  &   -- &  Border/background  \\ [5pt]	
	13 &   12:58:38& -17:22:48 &  4.6 & 0.032  &  -- &   Border/background  \\ [5pt]	
	14 & 12:56:34  & -17:43:40  &  4.4 & 0.033  &  -- &   Border/background  \\ [5pt]	
	15 & 12:57:50  & -17:10:20 &  4.3 & 0.029  &   -- &  background  \\ [5pt]	
	16 & 12:58:40  & -17:36:10 &  4.2 & 0.034  &  -- &   A1644 candidate  \\ [5pt]	
	17 & 12:57:48  & -17:32:42 &  4.2 & 0.029  &  -- &    Border/background  \\ [5pt]	
	18 & 12:57:50  & -17:05:52 &  4.1 & 0.028  &  -- &    Border/background  \\ 		
	
\hline
\hline
\end{tabular}
\end{center}
\end{table*}

\subsubsection{Correlation with cluster luminosity map}
\label{sec:rec.light}

The most straightforward way to identify which of the peaks in the convergence map belong to A1644 is to identify possible counterparts in the luminosity map. It assumes that mass and (optical) light will follow each other, which tends to hold even for merging clusters 
\citep{clowe04,clowe06}.

\begin{figure*}
\begin{center}
\includegraphics[width=\textwidth,angle=0]{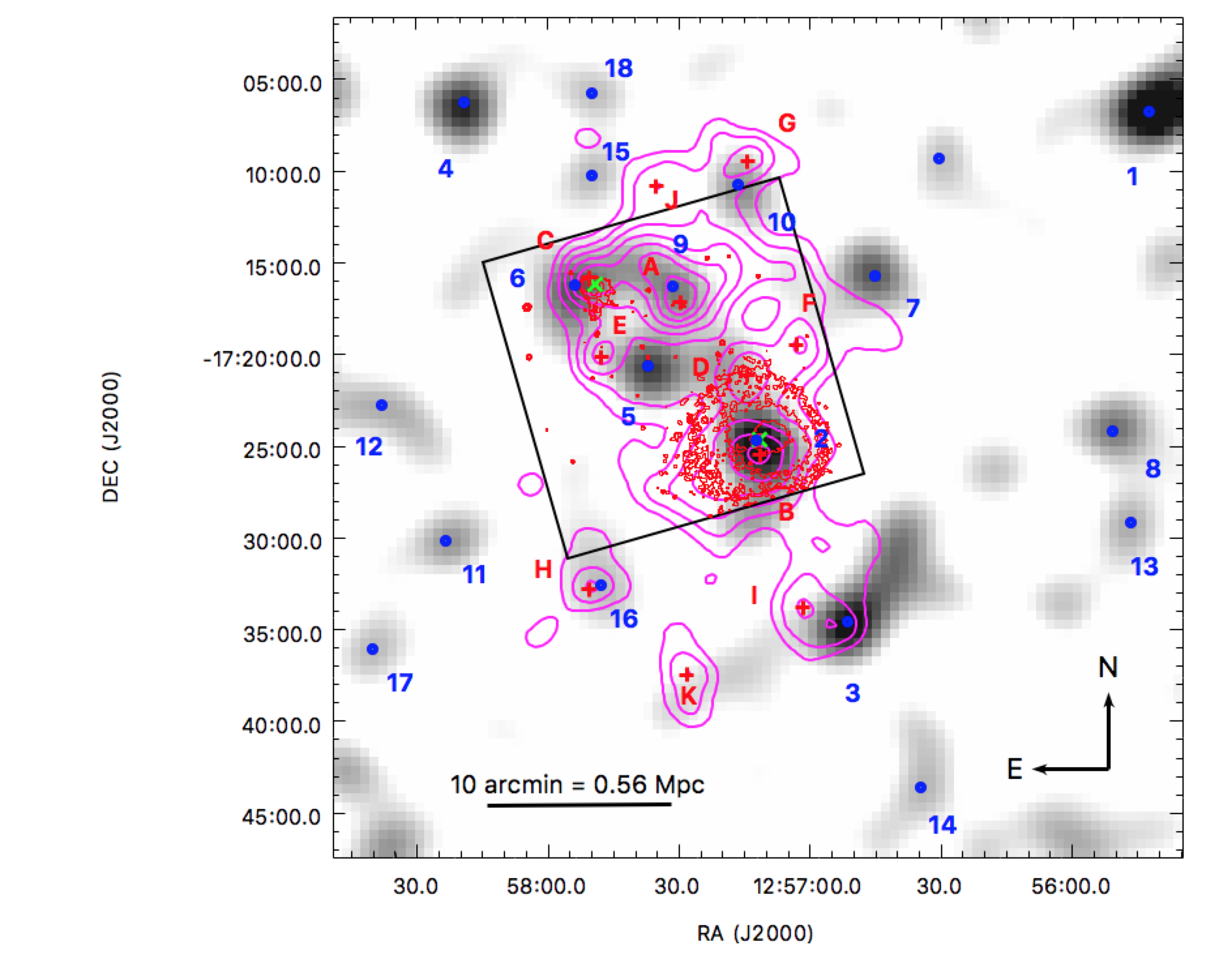}
\caption[]{Mass-luminosity correlation. Projected mass map (grey-scale) overlaid with the luminosity density distribution of red cluster members (magenta contours). Blue dots mark the mass peak centres (labelled 1--18) whereas the red ``+'' are placed at the galaxy clumps centre (A--H). BCGs positions are marked with green ``$\times$'' symbols. The black box shows the X-ray \textit{Chandra} image field of view.}
\label{fig:mass.red}
\end{center}
\end{figure*}

%<--Intro
We search for peaks in the red-sequence map in a similar manner than we did in the $\kappa$ map, this time with a circular window of 1.2 arcmin radius. We found 11 clumps (labelled A--K) with significance greater than $3.5\sigma$\footnote{This slightly lower threshold was chosen so that we could find viable counterparts for all $\kappa$ peaks.} ($\sigma$ being the standard deviation of all pixels). We also computed the nearest mass clump and the respective distance. These values are presented in Table~\ref{tab:red.seq.peaks}. In Fig.~\ref{fig:mass.red} we show the $r'$ projected galaxy luminosity distribution overlapped with the projected mass map.

%<--Table: red-sequence distribution
\begin{table}
\caption[]{Characteristics of the peaks found in the luminosity map of red-sequence galaxies. Columns are, respectively: ID, position, significance in units of $\sigma$ (standard deviation of all pixels), the nearest mass peak ID and the respective projected distance assuming A1644 redshift.}
\label{tab:red.seq.peaks}
\begin{center}
\begin{tabular}{c c c c c c}
\hline
\hline 
 Clump & $\alpha$  & $\delta$ & S & Nearest   & Distance\\
  ID & (J2000) & (J2000) & ($\sigma$)  & mass clump  &(kpc)\\
 
\hline
	A & 12:57:30  & -17:17:12 &  5.8 & \#9  & 51  \\[5pt] 
	B & 12:57:11  & -17:25:32 &  5.0 & \#2  & 42  \\[5pt]
	C & 12:57:50  & -17:15:51 &  4.9 & \#6  & 49  \\[5pt]
	D & 12:57:14  & -17:21:12 &  4.4 & \#2  & 201 \\[5pt]	
	E & 12:57:48  & -17:20:11 &  4.2 & \#5  & 145 \\[5pt]	
	F & 12:57:03  & -17:19:32 &  4.1 & \#7  & 314 \\[5pt]
	G & 12:57:14  & -17:09:32 &  3.8 & \#10 & 78 \\[5pt]	
	H & 12:57:51  & -17:32:51 &  3.6 & \#16 & 39 \\ [5pt]
	I & 12:57:02  & -17:33:52 &  3.6 & \#3  & 146 \\ [5pt]
	J & 12:57:35  & -17:10:52 &  3.5 & \#15 & 195 \\ [5pt]
	K & 12:57:28  & -17:37:32 &  3.5 & \#16 & 372 \\ 
\hline
\hline
\end{tabular}
\end{center}
\end{table}

%<--Figure: A1644 mass versus light (red-sequence)

From Fig. \ref{fig:mass.red} and Table~\ref{tab:red.seq.peaks} we can see that A1644S and A1644N1 are, as expected, noticeable. They are associated with the luminosity peaks B (\#2) and C (\#6), respectively, which offset of only 42 and 49 kpc\footnote{Projected distances assuming A1644 redshift.} between luminosity and projected density peaks.

The most prominent peak in the luminosity map (A), is the one we named A1644N2 in Sec. \ref{sec:red}. It is clearly associated with the mass density peak \#9, only 51 kpc (55~arcsec) away.  Due to the absence of any detectable X-ray emission from this region on the \textit{Chandra} image (Fig. \ref{fig:density}), A1644N2 can be characterized as gas-poor, if not gas-free. 

Despite the absence of a dominant red galaxy, as seen in A1644S and A1644N1, there is a large number of galaxies in A1644N2. The significance rank of luminosity presented in Table~\ref{tab:red.seq.peaks} does not change significantly when we vary the smoothing scale. For smaller scales (5, 10 and $15$~arcsec) clump A is subdivided into two others which became the first and second most significant. For larger scales (30 and 40 arcsec), clumps A and C are merged into a single and most significant clump.

%<--Other related structures
On a lower significance level, we can find two other A1644 substructures by associations between peaks in the red members luminosity and mass density maps: G-\#10 offset by 78 kpc) and H-\#16 (39 kpc). Both are on the outskirts of the A1644's core formed by S, N1 and N2 substructures. They are both outside the \textit{Chandra} image field of view.

All other galaxy concentrations are, at least, 145 kpc away from the nearest mass clump, about three times more than A-\#9, B-\#2 and C-\#6, and thus their physical associations cannot be taken for granted. E-\#5 is one of such cases. It is within the core of the cluster and is one of the highest peaks of the density mass map.  It is also well inside the \textit{Chandra} field of view and has no detectable X-ray emission. 

Another case is I-\#3. It resembles H-\#16 and G-\#10 by being outside the cluster core and the X-ray image as well. The projected distance between mass and light centres is indeed larger (146 kpc), but as it is in a much less crowded region of the field, their association seems plausible.

In Section \ref{sec:dynamical} we will use the radial velocity of cluster members in an attempt to solve, among other issues, the question about the pertinence or not of the aforementioned haloes to A1644.

%<--Possible background structures?

In order to check whether some of the mass map density peaks would be associated with some massive background structure, we looked for 
associations between them and overdensities in the background galaxy sample. We failed to find any. Our interpretation is, therefore, that most of the $\kappa$ peaks not associated with A1644 should be due to constellations of ($\lesssim 10^{13}$ M$_\odot$) background haloes which dominate the population of low peaks in the field \citep[e.g.][]{Yang13,Liu16,Wei18}.

\subsection{Mass estimation}
\label{sec:mod.mass}

\subsubsection{Mass field modelling}
\label{sec:modelling}

%<--What are the structures we are interested in?
%After the identification of the mass clumps related to A1644, 

The weak lensing masses of the individual mass clumps are in general better recovered by fitting physically motivated profiles. In crowded fields, such as the one we are dealing with, it can be done individually as there will be clear covariances between the several substructures. Here we will fit simultaneously 18 structures, which we will treat as dark matter haloes centred on the peaks with $\nu>4$ found in the $\kappa$ map.

%<--Mass modelling description
The weak gravitational lensing effect induced on each source galaxy is a combination of those generated by each of the individual mass clumps. Here, we will adopt a simplified model in which all of them are located at the same cluster redshift. It is easier, in this case, to deal with the Cartesian components of the reduced shear, $g_1$ and $g_2$, which are obtained by projecting the lens generated tangential component $g_+$ by the lensing convolution kernel,
\begin{equation}
D_1 = \frac{y^2 - x^2}{x^2 + y^2}\mbox{,}
\quad
D_2 = \frac{-2xy}{x^2 + y^2},\\
\label{A1644.eq.kernel}
\end{equation} 
where $x$ and $y$ are the Cartesian coordinates relative to the lens centre which was considered at the respective mass peak position.

The resulting reduced shear that will affect the image of each source galaxy can be written as
\begin{equation}
g_i = \sum_{k=1}^{N_{\rm clumps}}  g_i^{k},
\label{eq:g_sum_clumps}
\end{equation}
with $i\in\{1,2\}$. As we modelled the mass peaks in the region of interest, $N_{\rm clumps}=18$.

The model adopted is as a single circular NFW profile per clump or halo \citep[][]{nfw96,nfw97}. The $\chi^2$-statistic is then
\begin{equation}
\chi^2=\sum_{j=1}^{N_{{\rm sources}}} \sum_{i=1}^{2}  \frac{[g_i(M_{200},c)-e_{i,j}]^ 2}{\sigma_{\rm int}^2+\sigma_{{\rm obs}_{i,j}}^2}, \label{eq:chi2_d}
\end{equation}
where $g_i(M_{200},c)$ is the reduced shear (Eq.~\ref{eq:g_sum_clumps}) produced by all the NFW haloes over the position of a given source galaxy, $M_{200}$ is the mass within a radius where the mean enclosed density is equal to $200 \times$ the critical density of the Universe, $c$ is the halo concentration parameter, $e_{i,j}$ is the measured ellipticity parameter per galaxy, $\sigma_{{\rm obs}_{i,j}}$ is its measurement uncertainty and $\sigma_{\rm int}$ is the shape noise, taken as $\sim0.35$ for our data. 

The likelihood can now be written as
\begin{equation}
\ln \mathcal{L} \propto - \frac{\chi^2}{2}\mbox{.} 
\label{eq:likelihood.d}
\end{equation}
%

%<--Strategies to reduce the degrees of freedom
There are $2\times N_{\rm clumps}$ free parameters in our model which can lead to instabilities in the final model. We thus opt to reduce this number in half by adopting the $M_{200}-c$ relation by \citet{duffy08}: 
%<--Equation: Duffy relation (http://arxiv.org/pdf/0804.2486v4.pdf)
\begin{equation}
c=5.71\left(\frac{M_{200}}{2\times10^{12}h^{-1}M_{\odot}}\right)^{-0.084}(1+z)^{-0.47}.
\label{eq:duffy_rel}
\end{equation}
Given that the above relation having a scatter, we opt to not include it in our model in order to keep the parameter space to a minimum, given the complexity of the field.

Finally, the posterior can be written as
\begin{equation}
 {\rm Pr}( M_{200}|\rm{data}) \propto  \mathcal{L}(\rm{data}|M_{200})\times \mathcal{P}(M_{200})\mbox{,}
\label{eq:posterior}
\end{equation} 
where we adopted an uniform prior $\mathcal{P}(M_{200})$, $0<M_{200}\leq 8\times 10^{15}$ $M_{\odot}$. %aiming to to accelerate the model convergence.}

%<--Model assessment
\subsubsection{Model assessment}
\label{sec:mod.assessment}

%<--Description
Before stating the results, given the complexity of this field, mostly driven by the low signal from the very nearby cluster, we felt the need to further ensure the reliability of the whole modelling procedure.  We created a synthetic reduced shear field by positioning NFW lenses at the same position of the eighteen detected mass density peaks plus a population of source galaxies also at the same position as the ones in our background sample.

For this model, we got the mass of A1644S from the  X-ray mass by
\cite{Johnson10}\footnote{The value was converted from $M_{500}$ to $M_{200}=4.65\times10^{14}$ M$_\odot$ (following an NFW profile) to be used here.}. The remaining masses were set by multiplying this value to the ratio of the peak heights ($\nu / \nu_{A1644S}$; Table~\ref{tab:masses}). 
The reduced shear components for each galaxy were then set as the sum of the contributions of each of the lenses. We did not add anything else to take into account the shape or measurement noises. 

Given the high degree of idealisation idealization of this simulation, one should expect all the results to be very close to the true ones, and it happens for those with higher peak heights and on the core of the cluster.

%<--Results
However, in some cases, the recovered representative mass value\footnote{Corresponding to the median of each marginalized posterior.} was considerably larger than the reference ($>35\%$). We mention, as an example, the case of clump \#14 whose measured mass was 56\% larger than stated in the synthetic catalogue. These mass clumps have in common: (i) their relative low significance $\nu$ and/or (ii) their proximity to the border of the region considered for modelling. Based on these arguments, we decided not to consider the MCMC results for the clumps with $\nu<5.5$, nor those located close to the borders.

\subsubsection{Results}
\label{sec:results}

%<--Brief MCMC description
We restricted the source galaxies used in the model to those contained in the central region of interest (Fig.~\ref{fig:field}). 
Then, we sampled the posterior (Eq.~\ref{eq:posterior}) using the MCMC algorithm with a simple Metropolis sampler {\sc MCMCmetrop1R} \citep{MCMCpack} built in R environment. We generated four chains with $10^5$ iterations each one, removing the first $10^4$ as ``burn-in''. The final convergence, checked through the potential scale factor $R$ implemented in the {\sc Coda} library \citep{coda}, is attested within 68 per cent c.l. ($R\approx1$).

The results, median of the marginalized values of $M_{200}$, can be seen in Table~\ref{tab:masses}. All the posteriors are unimodal, although not symmetrical (and hence the difference in the plus and minus uncertainties) and not degenerate. The higher correlations are still small -- namely between the masses of A1644S and \#5 (-0.25) and A1644S and \#3 (-0.21).

As we can observe in Table~\ref{tab:masses} the ratio of fitted mass to mass uncertainty is smaller than the detection \citep[e.g.][]{Martinet16}. One possible mechanism responsible for that is the non-conformity of the model assumptions with the real clumps, for example, the absence of axisymmetry and/or a deviation from the theoretical profile.

%<--Obvious cluster-related substructures

The haloes directly related to the X-ray emission, A1644S (\#2) and A1644N1 (\#6), have masses of  $1.90_{-1.28}^{+0.89}\times 10^{14}$ M$_\odot$ and  $0.90_{-0.85}^{+0.45} \times10^{14}$ M$_\odot$, respectively. Both are considerably smaller than X-ray measurements presented by \cite{Johnson10} (A1644S: $4.6\pm0.6$; A1644N1: $3.8\pm0.6$). The discrepancy is noticeable but not a total surprise as the A1644 ICM is out of hydrodynamical equilibrium and this surely affects the X-ray mass estimation.

As for A1644N2 (\#9) we got  $M_{200} = 0.76_{-0.75}^{+0.37} \times10^{14}$ M$_\odot$. Therefore, the core of A1644 has a combined mass of  $3.99_{-1.59}^{+1.39} \times10^{14}$ M$_\odot$, corresponding to the sum of the  S+N1+N2 posteriors. Adding the candidate clump \#5, the value increases to  $5.11_{-1.44}^{+1.32}\times10^{14}$ M$_\odot$.

%<--Comments about \#10, #5 and #3.

Substructure \#3 came out of the analysis as the single most massive clump of the field, although with rather large error bars:  $3.37_{-2.06}^{+1.40}\times 10^{14}$ M$_\odot$. It was rather unexpected, given that its convergence peak is almost the same height as \#2 (A1644S), whose nominal mass value is almost half of it.  Also there are not so many cluster galaxies associated with it, assuming a I-\#3 connection, nor any hints of a background structure in the photometry data. From the $\kappa$ map (Fig. \ref{fig:mass.red}) it seems that it may have an extension or (projected) neighbours to the NW direction, which might enhance its mass somewhat.

Regarding the A1644 candidate structures, our model was not able to recover the mass of \#16. The masses of \#5 and \#10  are $0.87_{-0.82}^{+0.43}\times 10^{14}$ and  $0.56_{-0.55}^{+0.28}\times 10^{14}$ M$_\odot$, respectively.

\section{Dynamical analysis}
\label{sec:dynamical}

From the radial velocities available in the literature we explore the complex dynamical picture of A1644. We are particularly interested in verifying whether the mass substructures found through our photometric plus weak lensing analysis are actually dynamical structures bound to the main cluster.

\subsection{Identification of the spectroscopic cluster members}
\label{sec:spec.data}

\begin{figure}
 \begin{center}
\includegraphics[width=\columnwidth, angle=0]{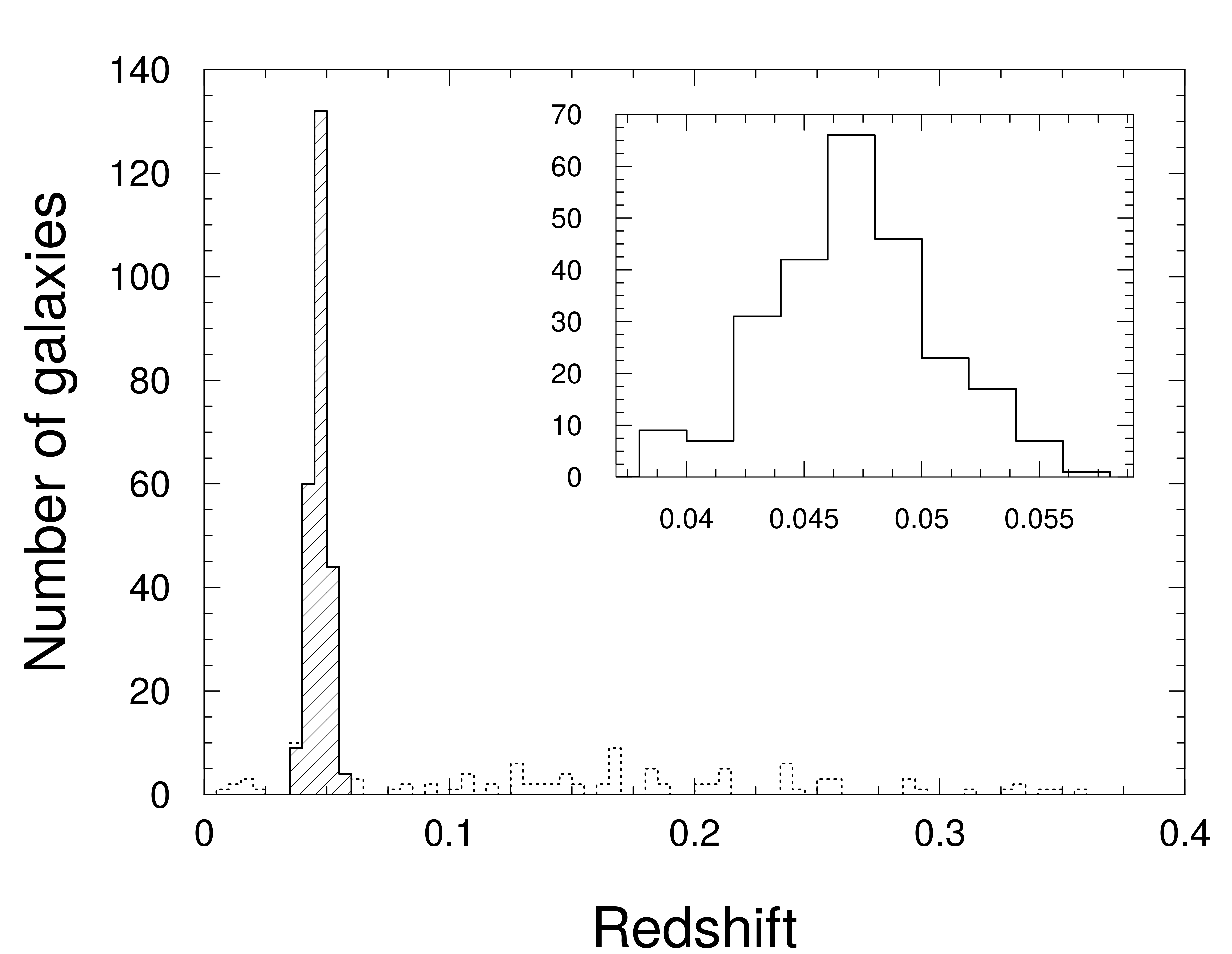}
%/home/monteiro/work/A1644/programs/R/velocities.R (@portoalegre)
\caption[]{Redshift distribution in A1644 region. We have selected the spectroscopic cluster members after a $3\sigma$-clipping procedure. The selected members are highlighted by hashing lines and are shown in more detail in the upper right panel. The 249 selected members  have $\bar{z}=0.0471\pm0.0002$ with $\sigma_v/(1+z)=1017$ km s$^{-1}$.} 
\label{fig:redshift}
\end{center}
\end{figure}

%<--Origin of the data
The field of A1644 has been exhaustively observed by the WIde-field Nearby Galaxy-cluster Survey \citep[WINGS;][]{Wings}. As a result, a simple search on the NASA Extragalactic Database (NED)\footnote{\texttt{https://ned.ipac.caltech.edu/}} has revealed a total of 360 galaxies with available redshifts located inside a circular region of 33 arcmin centred on A1644S. After searching on our galaxy catalogue, we have found photometric counterparts for 341 objects whose redshift distribution can be seen in Fig.~\ref{fig:redshift}.

%<--Figure: Redshift distribution

%<--Search for the spectroscopic members
A simple approach to select the cluster spectroscopic members consists of the application of the 3-$\sigma$ clipping method \citep{3sigmaclip}. Despite its simplicity, this procedure is sufficiently accurate to remove the most obvious outliers of the sample \citep{Wojtak07}. The resulting sample consisted of 249 galaxies with $\bar{z}=0.0471 \pm 0.0002$ and $\sigma_v/(1+\bar{z})=1017$ km s$^{-1}$\footnote{The $(1+z)^{-1}$ factors that appears multiplying velocity dispersions or differences are there to compensate for the cosmological stretching of velocities.} (hatched region in Fig.~\ref{fig:redshift}). According to the Anderson-Darling test, the null hypothesis of Gaussianity cannot be discarded for this sample within a confidence level of 95\% (${\rm p-value}=0.21$). This is in agreement with recent findings of \cite{Lopes18}.

%<--Searching for subtle substructures
Although the overall distribution of redshifts is Gaussian it does not necessarily mean that there are no substructures present. To search for those,
we have applied the well-known $\Delta$-test \citep{ds}, that looks for structures in both velocity space and projected position, and are described by the following equations:
%
%<--Equation: DS  test
\begin{equation}
 \delta_i=\left \{ \left ( \frac{N_{\rm nb}+1}{\sigma^2} \right ) [(\bar{v}_{l}-\bar{v})^2+(\sigma_l - \sigma)^2] \right \}^{1/2}
 \label{eq:ds}
\end{equation}
%
%<--Equation: DS  statistic
\begin{equation}
\Delta=\sum_{i=1}^{N} \delta_i.
 \label{eq:sumds}
\end{equation}
For a cluster hosting any substructures, \cite{ds} suggest $\Delta \ge N$ as the expected statistics. However, \cite{hou12} have proposed an improved index based on the ${\rm p-value}<0.01$ to identify a substructure \citep[see][for a more complete description of this statistic]{Monteiro-Oliveira17a}.

\begin{figure*}
 \begin{center}
\includegraphics[width=1.0\textwidth, angle=0]{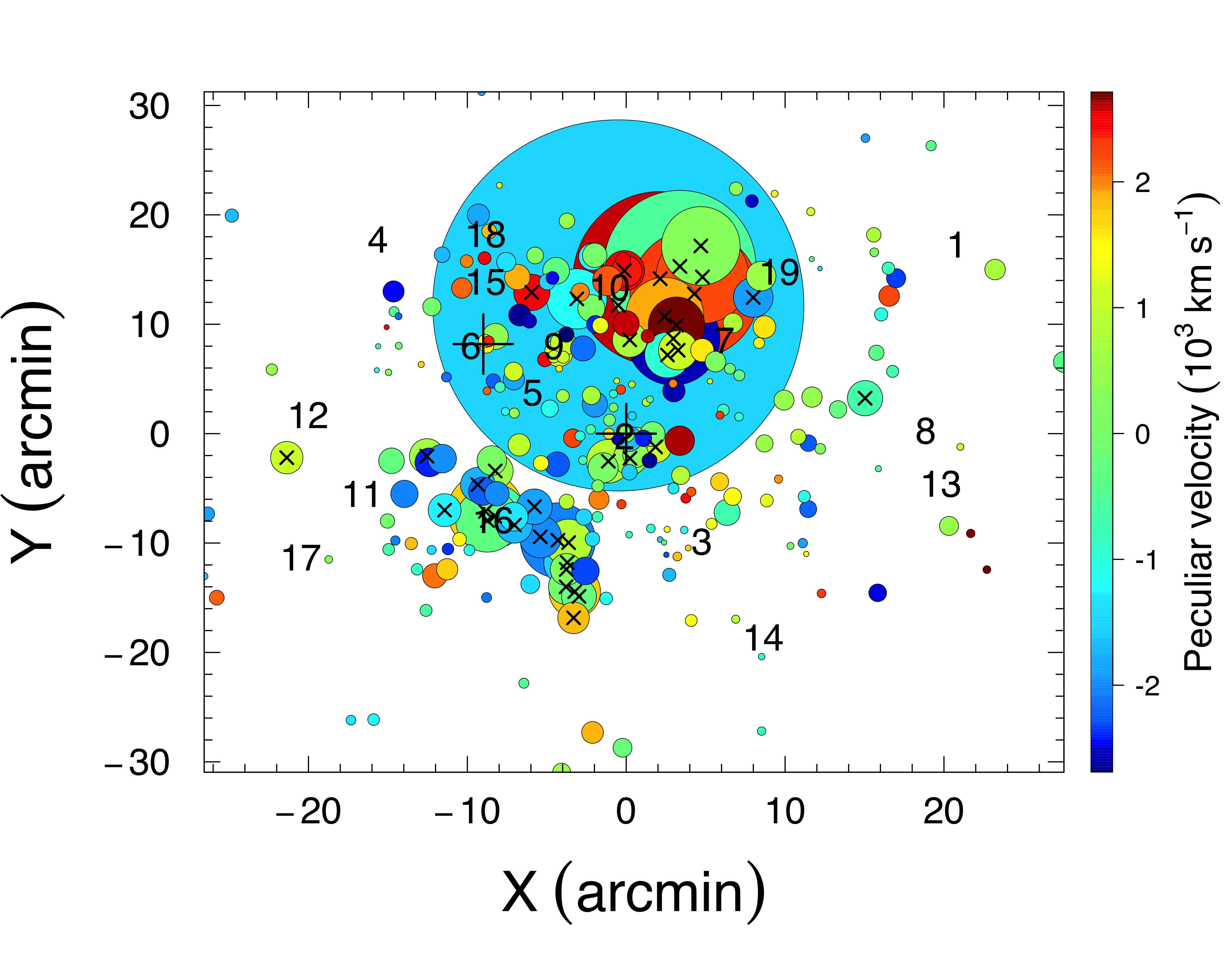} %/home/monteiro/work/A1644/programs/R/velocities.R (@portoalegre)
\caption[]{Bubble plot produced with $\Delta$-test results. The filled circles indicate each galaxy position and their radius are proportional to $e^{\delta_i}$ meaning that the larger this value, the more relevant the substructure is. To make their positions clear, two ``+'' signals are placed at the BCGs locations. The mass clump IDs are placed at the exact place of the corresponding centre. The colour bar represents the peculiar velocities $v_i-\bar{v}$ and ``$\times$'' indicates the galaxies classified as belonging to some substructure according to our $\delta_i \ge 2$ criterion. Our final sample consisted of 210 galaxies characterized by the same mean redshift and velocity dispersion of the initial sample. Note that this region is somewhat larger than those defined in Fig.~\ref{fig:field}.} 
\label{fig:ds}
\end{center}
\end{figure*}

%<--DS test results
The application of the $\Delta$-test in our sample returned $\Delta=339$ with a p-value of $0.007$, both pointing to the presence of substructures. The spatial distribution of the spectroscopic members overlaid with the $\Delta$-test results is shown as a ``bubble plot'' (Fig.~\ref{fig:ds}).

The $\Delta$-test does not identify which individual galaxies belong to a substructure. In \cite{Monteiro-Oliveira17a}, however, we envisaged a process to do so by searching for a threshold $\delta_c$ (equation~\ref{eq:ds}). All galaxies having $\delta_i\ge\delta_c$\footnote{We have adopted $\delta_c=2$ which has been shown to be a good threshold to remove the substructures.} were therefore considered part of some substructure and  then excluded of the main sample (galaxies marked with ``$\times$'' in Fig.~\ref{fig:ds}).

We ended up with 210 galaxies in the main sample with similar average and dispersion than the original 249 members sample. Most of the galaxies in substructures are spatially coincident with the mass clumps G-\#10 and H-\#16, previously identified as subcluster candidates. Thus, their high dynamical deviation with respect to the other members strongly suggest that both belong to a non-virialised regions being most probably infalling groups \citep{Einasto18}.

%<--Figure: DS test

\subsection{Identifying the subclusters with {\sc Mclust}}
\label{sec:Mclust}

%<--Introducing Mclust
The {\sc R}-based package {\sc Mclust} \citep{mclust} encompasses functions designed to perform multidimensional normal mixture modelling based on the analysis of finite mixture of distributions. It works by searching for the optimized clustering of the data using models with variable shape, orientation and volume. A prior restricting the number of normal components is allowed. The output includes number and characteristics of the components plus the individual memberships. More details about the use of {\sc Mclust} to search for substructures in the observable phase space of a cluster can be found in \cite{Einasto12}.

%<--1D investigation

At this point we already established that A1644 is composed of several substructures, probably in merger, but its  1D velocity distribution is very consistent with normalcy. It is a hint that the subclusters' motions are taking place almost parallel to the plane of the sky \citep[e.g.][]{Monteiro-Oliveira18, Wittman18a}. As a consequence, one-dimensional methods, i.e. those based only on the radial velocity information, will certainly fail to assign confidently the galaxies to a host subcluster. In fact, as a confirmation, {\sc 1D-Mclust}  pointed out that the single-component model is strongly favoured in relation to a bimodal distribution in the redshift space as revealed by Bayesian information criteria \citep[][]{kass95} as $\Delta{\rm BIC}=16$.

\begin{figure}
\begin{center}
\includegraphics[width=\columnwidth]{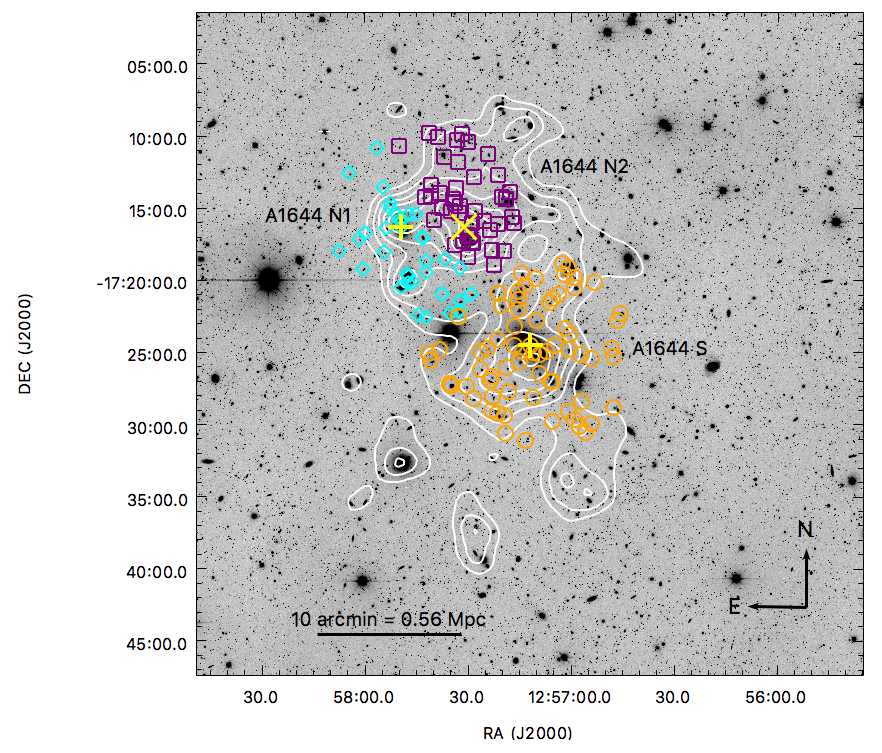}  
\caption[]{Galaxy classification suggested by double application of {\sc 2D-Mclust}. The picture shows the spatial distribution of the members of A1644S (orange), A1644N1 (cyan), and A1644N2 (purple) overlaid with the numerical density contours of the combined sample containing both spectroscopic and photometric members (white lines).  For comparison, ``+'' are placed at BCGs positions and ``$\times$'' marks the position of the mass clump A1644N2.}
\label{fig:A1644:dyn.ana}
\end{center}
\end{figure}

%<--Introducing and justifying the 2-D modelling
Given this scenario, we directed our efforts to the analysis of the spatial distribution of cluster members within the cluster core. The sample used here consists of the projected position ($\alpha$ and $\delta$) of combined photometric red-sequence and spectroscopic members. Since this data is spread across all the field, we have restricted our region of interest to the inner part, selecting only the galaxies inside two circular regions, each one with a radius of 7 arcmin, centred on the BCG S and on the mid-point between A1644N1 and A1644N2. See Fig. \ref{fig:A1644:dyn.ana} for the positions of the 163 galaxies of this sample.

%<--2D-Mclust results
The application of {\sc 2D-Mclust} showed the data are best described as a bimodal distribution. This model is strongly favoured with respect to the second-best model with three groups, as pointed out by $\Delta{\rm BIC}=12$.  In both scenarios the two main groups contain most of the sample. In the case with three groups, the third one has a few galaxies on the SE direction of A1644S (or clump B-\#1; Fig.~\ref{fig:mass.red}). The outcome is a cut in the sample along the SE--NW direction, separating A1644S at one side and  A1644N1/N2 at the other (orange points versus cyan and purple on  Fig. \ref{fig:A1644:dyn.ana}). This division is consistent with the  position of the most significant galaxy clumps according to Table~\ref{tab:red.seq.peaks}. We also performed the exercise to include the redshift information\footnote{For those galaxies for which this information is available.} for a tri-dimensional analysis with  {\sc 3D-Mclust}. The result was a return to unimodality.

%<--Figure: 2D tests results

In order to understand the dynamical process that formed the complex A1644 as we see today, we are interested in estimating the relative velocities between the clumps we found both by photometry and weak lensing.
In order to do that, we reran {\sc 2D-Mclust} over the northern sample it already identified and imposed a division in two groups. These results are presented in Fig.~\ref{fig:A1644:dyn.ana} (cyan versus purple points) and now we have clear samples that can be associated with each of the three main core structures (S, N1, and N2). 

In Table~\ref{tab:2dmclust}, we show redshifts and velocity dispersions from the redshifts we have for each of those samples. In Table~\ref{tab:delta.v}, we show the differences in (rest-frame) velocities. As expected we find small separations between them, in the velocity space, not inconsistent with zero given the uncertainties.

%<--Table: 2D tests results
\begin{table*}
\begin{center}
\caption[]{Statistics of galaxy groups found with {\sc 2D-Mclust}.}
\begin{tabular}{lccc}
\hline
\hline
	& A1644S & A1644N1 & A1644N2\\
\hline
Number of galaxies  	   & $42$	     & $17$  & $17$\\
$\bar{z}$  & $0.0472\pm0.0006$ & $0.0476\pm0.0007$  & $0.0460\pm0.0012$\\
$\sigma_v/ (1+\bar{z})$ (\kms) &  $1146_{-59}^{+211}$   &$886_{-85}^{+261}$ & $1376_{-134}^{+402}$\\
\hline
\hline
\end{tabular}
\label{tab:2dmclust}
\end{center}
\end{table*}

%<--Table: Separation in LOS
\begin{table}
\begin{center}
\caption[]{Dynamics of the subclusters A1644S, A1644N1, and A1644N2 modelled by {\sc 2D-Mclust}. Projected distances were computed based on the respective mass peak positions. All error bars correspond to 68 per cent c.l.}

\begin{tabular}{lcc}
\hline
\hline
Pairs	& $\delta v/(1+\bar{z})$ (km s$^{-1}$) & Projected distance (kpc) \\
\hline
S--N1  & $137\pm278$ & 723 \\
S--N2  & $334\pm378$ & 531 \\
N1--N2 & $471\pm397$ & 298 \\
\hline
\hline
\end{tabular}
\label{tab:delta.v}
\end{center}
\end{table}

\section{Hydrodynamical simulations}
\label{sec:hidro}

In this section, we employ $N$-body hydrodynamical simulations to evaluate whether the newly discovered structure A1644N2 could be responsible for the collision that gave rise to the sloshing spiral. Several simulations were performed in search of a model that reproduces some of the desired morphological features of A1644. Here we present one model that successfully recovers the spiral morphology of A1644S: it is an encounter between A1644S and A1644N2 only. In the simulation presented here, A1644N1 is not included. For a more detailed analysis of the set of simulations, the reader is referred to an accompanying paper \citep{Doubrawa2019}, in which we explore the outcomes of alternative scenarios and discuss their relative advantages and disadvantages. In this section, we will describe the main features of one plausible model.

To have an acceptable scenario, some constraints must be satisfied by the simulations: the virial masses and virial radii estimated in this paper; the spiral morphology of cold gas, with an extent of approximately 200\,kpc; the observed projected separation of $\sim$550\,kpc between the structures; and simultaneously A1644N2 must have a low fraction of gas to be nearly undetectable in X-ray observations.

% initial conditions
Taking the new virial mass estimates, $1.9 \times 10^{14}$ M$_{\odot}$ for A1644S and  $0.76 \times 10^{14}$ M$_{\odot}$ for A1644N2, we create two spherically symmetric galaxy clusters that are initially in hydrostatic equilibrium. The method for generating initial conditions is similar to those used in \cite{Machado2015}. Each cluster is created with $2 \times 10^6$ particles, divided equally between gas and dark matter, following the procedure described in \cite{Machado2013}. Simulations were carried out with {\sc Gadget-2} \citep{Springel2005}. Initially the clusters are separated by $3$ Mpc along the $x$-axis, with an impact parameter of $b = 800$ kpc, and initial relative velocity of $-700$ km s$^{-1}$ in the $x$ direction (at $t=0$). The evolution of the system is followed by $5$ Gyr.

\begin{figure}
\includegraphics[width=\columnwidth]{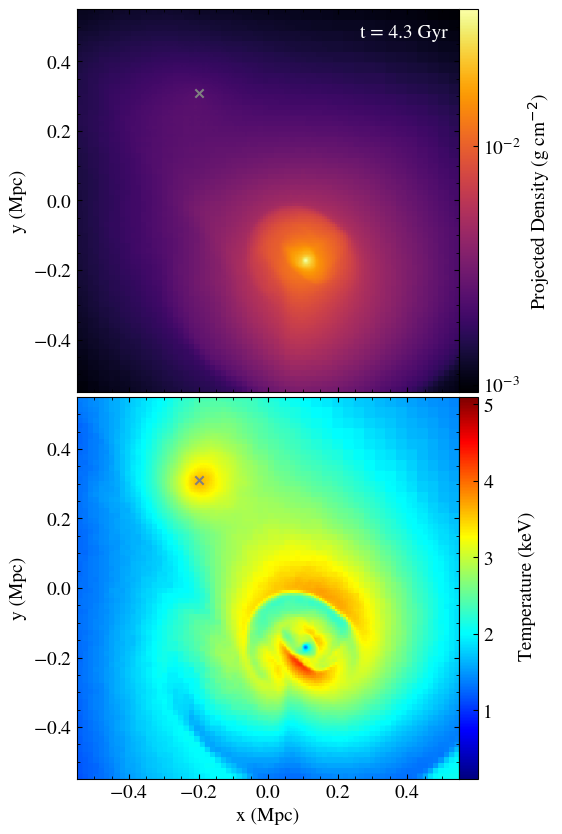}
\caption{Top panel: Projected density map for the merger simulation between A1644S and A1644N2, at $t=4.3$ Gyr. Bottom panel: Projected emission-weighted temperature. The symbol marks the presence of A1644N2, nearly undetectable on the density map.}
\label{Fig1}
\end{figure}

% dynamical results
The desired configuration is achieved at $t=4.3$ Gyr, that is, $1.6$ Gyr after the pericentric passage and $0.5$ Gyr after reaching the maximum separation of $770$ kpc.  
This model presents temperature maps that are in good qualitative agreement with the expected ranges presented in \cite{Reiprich2004} and \cite{Johnson10}. Similarly, the spiral morphology of cool gas in A1644S displays adequate shape and extent. Fig. \ref{Fig1} shows the projected density, and the emission-weighted temperature map of the hydrodynamical simulation at $t=4.3$ Gyr, highlighting the presence of A1644N2 with the symbol $\times$.  

After the first pericentric passage, A1644N2 loses almost all of its gas, making it difficult to identify in X-ray observations. It retains about $0.1-1$ per cent of its initial gas mass. In order to obtain a projected separation of approximately $550$ kpc between clusters, at the best match of morphology, the orbital plane was inclined by $i \approx 30 ^\circ$ with respect to the plane of the sky. This is compatible with the low line-of-sight velocity found by the observational analysis (Table \ref{tab:delta.v}).
We can estimate this relative velocity between A1644S and A1644N2 through the displacement between the clusters cores in the $z$-axis inside a short time step around $t=4.3$ Gyr. This approach results in a line-of-sight velocity of $\sim 100 \textrm{ km~s}^{-1}$, in agreement with the low values expected in a merging system near their maximum separation.

% discussion

Our goal was to recover the morphological features observed in A1644 through hydrodynamical simulations. To judge the adequacy of the simulation results, most comparisons were made by qualitative visual inspection of the morphology. The model presented is one of the possible combinations of the collision parameters. With this scenario, we were able to reproduce, in a best instant, several of the simulation constraints, such as morphology, temperature and extent of the sloshed gas, projected separation and low X-ray emission of A1644N2, with the virial masses and radii derived from the gravitational weak lensing analysis. Our current plausible model is thus the scenario in which A1644N2 is the disturber that induced the sloshing. Other scenarios involving A1644N1 instead of A1644N2 did not give rise to results as satisfactory as the model presented here. Alternative simulations are explored in more detail in \cite{Doubrawa2019}.

\section{Discussion}
\label{sec:discussion}

%<--What is new here?
We presented here a comprehensive analysis of the nearby merging galaxy cluster Abell 1644. With a combination of observational and computational techniques, we were able to propose a new description for the preceding collision events which led to the current state. We put forth a scenario in which the remarkable spiral-like structure seen in the X-ray emission of A1644S arose as a result of the interaction with the newly discovered subcluster A1644N2 and not due to A1644N1 as suggested by previous studies.

\begin{figure}
\includegraphics[width=\columnwidth]{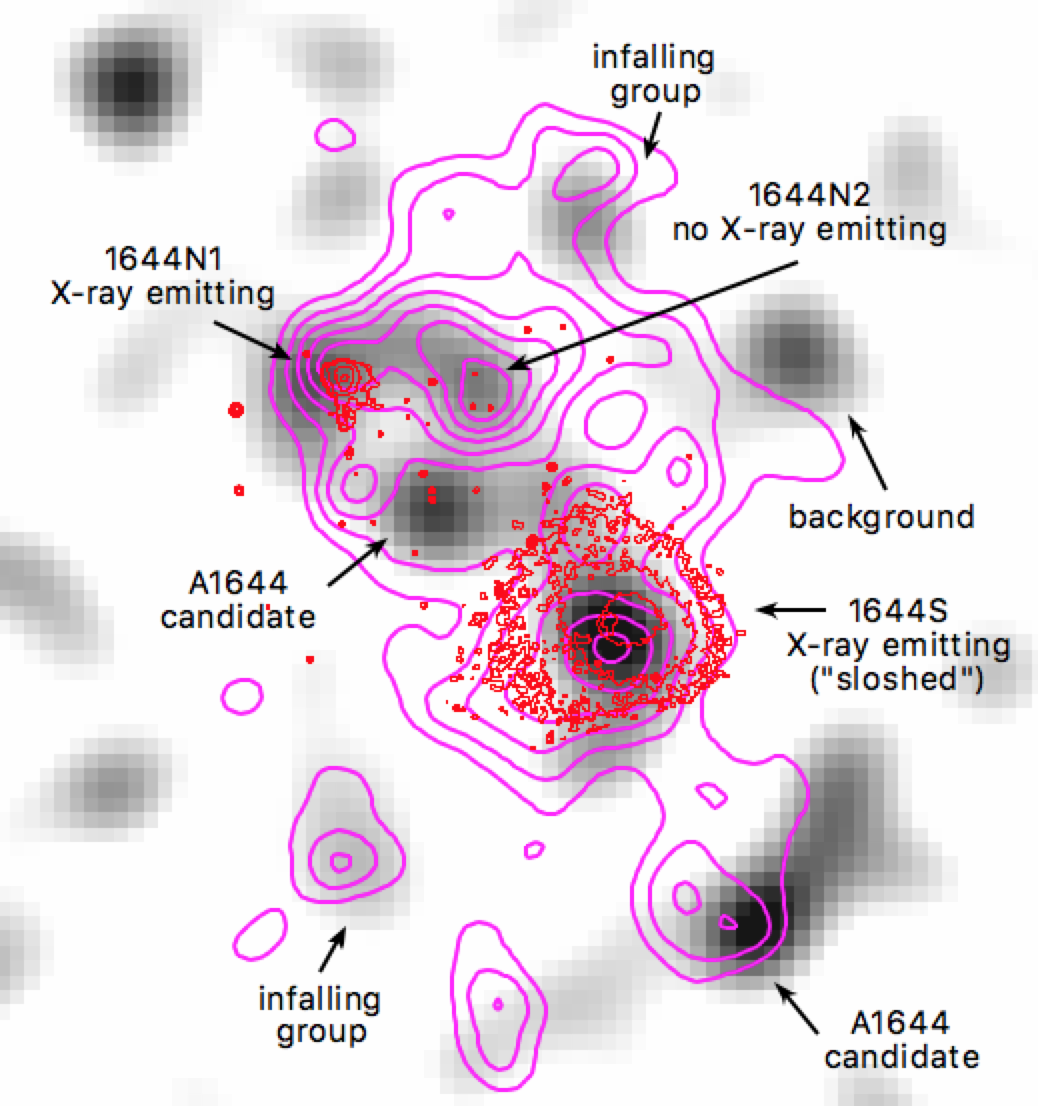}
\caption{Proposed scenario for the merging cluster A1644. The weak lensing mass map (gray-scale) is overlaid with the $r'$ projected galaxy luminosity distribution (magenta contours) and X-ray emission (red contours). The main subcluster, A1644S has a spiral-like pattern in its ICM distribution caused by the passage the gas-poor subcluster A1644N2, found in this work. The cluster core is also formed by the X-ray emitting subcluster A1644N1. There are also two infalling galaxy groups in the cluster vicinity. We classified as A1644 candidates two significant mass clumps surrounded by red galaxies.}
\label{fig:summary}
\end{figure}

%<--red-sequence spatial distribution
From large field-of-view images taken in three broad-bands, we built a careful selection of the galaxy samples. The luminosity map of the red cluster sequence galaxies (Fig.~\ref{fig:density}) brought the first hint that the cluster morphology should be somewhat more complex than previously claimed. Besides the two prominent galaxy clumps related to each BCG (A1644S and A1644N1), a third one, not reported in previous studies, was found slightly to the West of BGC N (A1644N2). Surrounding these central concentrations, two others can be found in the cluster outskirts. Further analysis based on their radial velocities, showed that they consist of infalling groups. There are also two galaxy clumps located reasonably close to significant mass concentrations, but with no spatial coincidence. With our analysis, we cannot categorically state if they are part of the cluster, though it could be possible. Conservatively, we classified both as A1644 candidates. However, for the description of the merger process among the main subclusters, the presence of such structures can be disregarded as a first approximation. These findings are illustrated in Fig.\ref{fig:summary}.

%<--Describing the mass distribution
In this work, we presented the first weak-gravitational lensing recovered projected mass distribution of the extremely low-$z$ merging cluster Abell 1644.  Since this technique does not assume any prior about the dynamical state, it is a powerful tool to recover the total mass of cluster haloes. The projected mass map corroborates the scenario indicated by the projected galaxy luminosity distribution, with each of the three main galaxy concentrations related to a respective mass clump. In A1644S, the BCG matches exactly the mass clump centre whereas A1644N1 is displaced only 56~kpc in relation to that. However, projected separations up to $\sim100$ kpc are not statistically significant because is also comparable with the expected effect caused by shape noise and smoothing of weak lensing mass maps \citep{dietrich12}. The three main structures are separated from each other by 723~kpc (S--N1), 531~kpc (S--N2), and 298~kpc (N1--N2).  The total mass content of A1644 is evaluated in, at least, $M_{200}=3.99_{-1.59}^{+1.39}\times 10^{14}$ M$_{\odot}$.

%<--Dynamical analysis
From the catalogue of radial velocities available in the literature, we were able to provide some insights about the dynamics of A1644. Regarding the three main subclusters, we found that they are not so apart in relation to the line of sight, having $
\delta v / (1+\bar{z})< 470$ km s$^{-1}$ (Table~\ref{tab:delta.v}) and, within the uncertainties, compatible with a negligible separation \citep[e.g.][]{Wittman18a}.

%<--Boost factor
One probe of the perturbed state of a system is how its velocity dispersion was boosted during the merger process \citep[e.g.][]{pinkney,Takizawa10,Monteiro-Oliveira18}. For an idealized isolated system,  the pre-merger velocity dispersion $\sigma_{\rm pre}$ can be estimated from an empirical relation $M_{200}$--$\sigma_v$ \citep[e.g.][]{biviano06}. To do this, we considered the entire sample of $4\times10^5$ mass values from the MCMC mass modelling (Sec.~\ref{sec:results}). We found $656_{-109}^{+146}$ km s$^{-1}$ for A1644S, $528_{-107}^{+135}$ km s$^{-1}$ for A1644N1 and  $502_{-101}^{+132}$ km s$^{-1}$ for A1644N2. After the pericentric passage, the velocity dispersion $\sigma_{\rm obs}$ is easily obtained (see Table~\ref{tab:2dmclust}). Therefore, the boost factors  $f\equiv\sigma_{\rm obs}/\sigma_{\rm pre}$, are $f_{\rm S}=1.8^{+0.4}_{-0.6}$, $f_{\rm N1}=1.7^{+0.6}_{-0.7}$ and $f_{\rm N2}=2.8^{+0.9}_{-1.2}$, respectively for A1644S, A1644N1 and A1644N2. The amount of events where the final velocity dispersion was not greater than unity (i.e. $\sigma_{\rm pre}$ was not enhanced by the collision)  was $2$ per cent, $10$ per cent and $2$ per cent  for A1644S, A1644N1, and A1644N2, respectively. However, these results should be considered with parsimony since velocity dispersion, specially from the Northern subclusters, was obtained with a small subset of galaxies (see Table~\ref{tab:2dmclust}).

%<--About A1644N2
The main finding of this work is the discovery of A1644N2. This structure presumably has a lower gas content than what would be expected for its $M_{200}$, since the X-ray emission map only shows gas counterparts for A1644S and A1644N1. At the position of A1644N2, the X-ray emission is not nearly as important as the others, even though its virial mass is comparable is those of A1644N1. What led to this situation?

One possible explanation is that A1644N2 suffered a recent close encounter with A1644S. If the pericentric passage was sufficiently close, A1644N2 could have been stripped of most of its gas. At the same time, its gravitational perturbation could have stirred the cool gas in the core of A1644S. This explanation has the advantage of simultaneously accounting for the two noteworthy gas features of this system: the sloshing spiral in the South and low gas content in the North. The intensity of the gas stripping in A1644N2 will depend on its precise orbit around A1644S and also on their initial gas density profiles. Similarly, the shape, size and temperature of the sloshing spiral depend on the density profiles and also on the parameters of the collision. The spiral morphology is sensitively time dependent, and so is the projected separation between the clusters. The question then would be: is there a combination of parameters capable of producing all of the desired outcomes at the same time? Our simulations provided one specific model in good qualitative agreement with the observational constraints. As is always the case in this type of modelling, one cannot ensure the uniqueness of the proposed solution. Nevertheless, the fact that a detailed model can be found indicates that a collision between A1644S and A1644N2 is indeed a feasible scenario. Other combinations of initial condition parameters are explored in greater detail in the associated paper \citep{Doubrawa2019}.

\section{Summary}
\label{sec:summary}

Here, we summarize the main findings of this work.

\begin{itemize}
    \item From our weak lensing analysis, based on deep and large field-of-view images, we have found that the merging galaxy cluster Abell 1644 ($z=0.0471$) presents a more complex structure than described in previous works. We point here that the system is comprised not only by two but by three central subclusters, surrounded by infalling galaxy groups.
    
    \item Subclusters A1644S and A1644N1 show each one a counterpart in the hot emitting X-ray gas. Their masses are, $M_{200}^{\rm S}=1.90_{-1.28}^{+0.89}\times10^{14}$  M$_\odot$ and $M_{200}^{\rm N1}=0.90_{-0.85}^{+0.45}\times10^{14}$  M$_\odot$, respectively. The velocity dispersions of their galactic contents are considerably high, $\sigma_v^{\rm S}=1146_{-59}^{+211}$ km s$^{-1}$ and  $\sigma_v^{\rm N2}=886_{-85}^{+261}$ km s$^{-1}$, respectively.

    \item The new structure identified in this work, A1644N2, is a gas poor subcluster, whose estimated total mass is $M_{200}^{\rm N2}=0.76_{-0.75}^{+0.37}\times10^{14}$  M$_\odot$. The velocity dispersion is evaluated to be $\sigma_v^{\rm N2}=1376_{-134}^{+402}$ km s$^{-1}$ based on its galaxy spectroscopic members.
    
    \item Combining the total mass distribution with the redshift catalogue of A1644 members, we identified two infalling groups at the cluster outskirts. The most massive has a mass of $M_{200}=0.56_{-0.55}^{+0.28}\times10^{14}$  M$_\odot$.
    
    \item We performed a set of tailored hydrodynamical simulations in which an encounter between A1644S and A1644N2 was the responsible for the rise of the sloshing spiral. We offered a specific model that accounts for several observed features, such as virial masses and radii, spiral morphology and temperature, separation and inclination. Our best model indicates that the collision is seen $1.6$ Gyr after pericentric passage and that the plane of the orbit is inclined with $30^\circ$ with respect to the plane of the sky. In this scenario, the two clusters have already reached the apoapsis and are currently incoming for their second approach.
    
\end{itemize}

Finally, we note that the full details of the set of hydrodynamical simulations are described in the companion paper \citep{Doubrawa2019}, where we explore alternative scenarios and also investigate what role A1644N1 might have played in the dynamics of this complex system.

\section*{Acknowledgements}
% Entry for the table of contents, for this guide only
\addcontentsline{toc}{section}{Acknowledgements}
We thank the anonymous referee for his/her suggestions that contributed to the improvement of the paper. This work has made use of the computing facilities of the Laboratory of Astroinformatics (IAG/USP, NAT/Unicsul), whose purchase was made possible by the Brazilian agency FAPESP (grant 2009/54006-4) and the INCT-A. Simulations were carried out in part at the Centro de Computa\c c\~ao Cient\'ifica e Tecnol\'ogica (UTFPR). This study was financed in part by the \textit{Coordena\c{c}\~{a}o de Aperfei\c{c}oamento de Pessoal de N\'ivel Superior - Brasil} (CAPES) - Finance Code 001.
RMO acknowledges support from CAPES.
REGM acknowledges support from the Brazilian agency \textit{Conselho Nacional de Desenvolvimento Cient\'ifico e Tecnol\'ogico} (CNPq) through grants 303426/2018-7 and 406908/2018-4.
GBLN thanks financial support from CNPq and FAPESP (grant 2018/17543-0).
ESC acknowledges support from the CNPq through grant 308539/2018-4
This work is based in part on data products produced at Terapix available at the Canadian Astronomy Data Centre as part of the Canada-France-Hawaii Telescope Legacy Survey, a collaborative project of NRC and CNRS.
We made use of the NASA/IPAC Extragalactic Database, which is operated by the Jet Propulsion Laboratory, California Institute of Technology, under contract with NASA.
This research draws upon DECam data as distributed by the Science Data Archive at NOAO. NOAO is operated by the Association of Universities for Research in Astronomy (AURA) under a cooperative agreement with the National Science Foundation. 

%%%%%%%%%%%%%%%%%%%%%%%%%%%%%%%%%%%%%%%%%%%%%%%%%%



\bibliographystyle{mnras}
\bibliography{monteiro-oliveira_library}

%%%%%%%%%%%%%%%%%%%%%%%%%%%%%%%%%%%%%%%%%%%%%%%%%%

%%%%%%%%%%%%%%%%% APPENDICES %%%%%%%%%%%%%%%%%%%%%
%\onecolumn

%%%%%%%%%%%%%%%%%%%%%%%%%%%%%%%%%%%%%%%%%%%%%%%%%%

% Don't change these lines
\bsp	% typesetting comment
\label{lastpage}
\end{document}